%% file: SMP-25-001_temp.tex
\documentclass[11pt,twoside,a4paper,cmspaper,final,collab]{cms-tdr}
\def\svnVersion{ba5e28d-D}\def\svnDate{2026/08/06}\def\cmsCernNoTag{CERN-EP-2026-210}\def\cmsCernDate{\today}\def\cmsMessage{Submitted to Physics Letters B}
\begin{document}\cmsNoteHeader{SMP-25-001}

\newcommand{\DeepJet}{{\textsc{DeepJet}}\xspace}
\newcommand{\PHiggs}{\ensuremath{\PH}}
\newcommand{\msd}{\ensuremath{m_{\text{SD}}^{\text{jet}}}\xspace}
\newcommand{\mwv}{\ensuremath{m_{\PW\PV}}\xspace}
\newcommand{\mWV}{\ensuremath{m_{\PW\PV}}\xspace}
\newcommand{\WV}{\ensuremath{\PW\PV}\xspace}
\newcommand{\Wj}{\ensuremath{\PW{+\text{jets}}}\xspace}
\newcommand{\Wlep}{\ensuremath{\PW^{\Pell}}}
\newcommand{\lint}{138\fbinv}
\renewcommand{\ie}{{i.e.}\xspace} 
\newcommand{\WW}{\ensuremath{\PW\PW}\xspace}
\newcommand{\WZ}{\ensuremath{\PW\PZ}\xspace}
\newcommand{\chwb}{\ensuremath{c_{\PH\PW\PB}}}
\newcommand{\chwbtil}{\ensuremath{c_{\PH\widetilde{\PW}\PB}}}
\newcommand{\chu}{\ensuremath{c_{\PH\PQu}}}
\newcommand{\chd}{\ensuremath{c_{\PH\PQd}}}
\newcommand{\chj}[1]{\ensuremath{c_{\PH\PQq}^{(#1)}}}
\newcommand{\cw}{\ensuremath{c_{\PW}}}
\newcommand{\cwtil}{\ensuremath{c_{\widetilde\PW}}}
\newcommand{\clj}[1]{\ensuremath{c_{\Pl\PQq}^{(#1)}}}
\newcommand{\clu}{\ensuremath{c_{\Pl\PQu}}}
\newcommand{\cld}{\ensuremath{c_{\Pl\PQd}}}
\newcommand{\cjj}[2]{\ensuremath{c_{\PQq\PQq}^{(#1,#2)}}}
\newcommand{\cju}[1]{\ensuremath{c_{\PQq\PQu}^{(#1)}}}
\newcommand{\cjd}[1]{\ensuremath{c_{\PQq\PQd}^{(#1)}}}
\newcommand{\cqj}[2]{\ensuremath{c_{\PQQ\PQq}^{(#1,#2)}}}
\newcommand{\Qw}{\ensuremath{\mathcal{Q}_{\mathrm{W}}}}
\newcommand{\QwDef}{\ensuremath{\varepsilon^{ijk} W_{\mu}^{i\nu} W_{\nu}^{j\rho} W_{\rho}^{k\mu}}}
\newcommand{\Qwtil}{\ensuremath{\mathcal{Q}_{\mathrm{\widetilde W}}}}
\newcommand{\QwtilDef}{\ensuremath{\varepsilon^{ijk} {\widetilde W}_{\mu}^{i\nu} W_{\nu}^{j\rho} W_{\rho}^{k\mu}}}
\newcommand{\Dbid}{\ensuremath{{\overset\leftrightarrow{D_{\mu}}}}}
\newcommand{\Qhwb}{\ensuremath{\mathcal{Q}_{\mathrm{HWB}}}}
\newcommand{\QhwbDef}{\ensuremath{H^{\dagger}H\sigma^{i} W_{\mu\nu}^{i}B^{\mu\nu}}}
\newcommand{\Qhwbtil}{\ensuremath{\mathcal{Q}_{\mathrm{H{\widetilde W}B}}}}
\newcommand{\QhwbtilDef}{\ensuremath{H^{\dagger}\sigma^{i}H {\widetilde W}_{\mu\nu}^{i}B^{\mu\nu}}}

\newcommand{\QhjOne}{\ensuremath{\mathcal{Q}_{\mathrm{Hq}}^{(1)}}}
\newcommand{\QhjOneDef}{\ensuremath{(H^{\dagger}i\overset\leftrightarrow{D_{\mu}}H)(\overline{q}\gamma^{\mu}q)}}
\newcommand{\QhjThree}{\ensuremath{\mathcal{Q}_{\mathrm{Hq}}^{(3)}}}
\newcommand{\QhjThreeDef}{\ensuremath{(H^{\dagger}i\overset\leftrightarrow{D_{\mu}^{i}}H)(\overline{q}\sigma^{i}\gamma^{\mu}q)}}
\newcommand{\Qhu}{\ensuremath{\mathcal{Q}_{\mathrm{Hu}}}}
\newcommand{\QhuDef}{\ensuremath{(H^{\dagger}i\overset\leftrightarrow{D_{\mu}}H)(\overline{u}\gamma^{\mu}u)}}
\newcommand{\Qhd}{\ensuremath{\mathcal{Q}_{\mathrm{Hd}}}}
\newcommand{\QhdDef}{\ensuremath{(H^{\dagger}i\overset\leftrightarrow{D_{\mu}}H)(\overline{d}\gamma^{\mu}d)}}

\newcommand{\QljOne}{\ensuremath{\mathcal{Q}_{\mathrm{lq}}^{(1)}}}
\newcommand{\QljOneDef}{\ensuremath{(\overline{l}_{p}\gamma_{\mu}l_{r})(\overline{q}\gamma^{\mu}q)}}
\newcommand{\QljThree}{\ensuremath{\mathcal{Q}_{\mathrm{lq}}^{(3)}}}
\newcommand{\QljThreeDef}{\ensuremath{(\overline{l}_{p}\sigma^{i}\gamma_{\mu}l_{r})(\overline{q}\sigma^{i}\gamma^{\mu}q)}}
\newcommand{\QjjOneOne}{\ensuremath{\mathcal{Q}_{\mathrm{qq}}^{(1,1)}}}
\newcommand{\QjjOneOneDef}{\ensuremath{(\overline{q}\gamma_{\mu}q)(\overline{q}\gamma^{\mu}q)}}
\newcommand{\QjjOneEight}{\ensuremath{\mathcal{Q}_{\mathrm{qq}}^{(1,8)}}}
\newcommand{\QjjOneEightDef}{\ensuremath{(\overline{q}T^{a}\gamma_{\mu}q)(\overline{q}T^{a}\gamma^{\mu}q)}}
\newcommand{\QjjThreeOne}{\ensuremath{\mathcal{Q}_{\mathrm{qq}}^{(3,1)}}}
\newcommand{\QjjThreeOneDef}{\ensuremath{(\overline{q}\sigma^{i}\gamma_{\mu}q)(\overline{q}\sigma^{i}\gamma^{\mu}q)}}
\newcommand{\QjjThreeEight}{\ensuremath{\mathcal{Q}_{\mathrm{qq}}^{(3,8)}}}
\newcommand{\QjjThreeEightDef}{\ensuremath{(\overline{q}\sigma^{i}T^{a}\gamma_{\mu}q)(\overline{q}\sigma^{i}T^{a}\gamma^{\mu}q)}}
\newcommand{\Qlu}{\ensuremath{\mathcal{Q}_{\mathrm{lu}}}}
\newcommand{\QluDef}{\ensuremath{(\overline{l}_{p}\gamma_{\mu}l_{r})(\overline{u}\gamma^{\mu}u)}}
\newcommand{\Qld}{\ensuremath{\mathcal{Q}_{\mathrm{ld}}}}
\newcommand{\QldDef}{\ensuremath{(\overline{l}_{p}\gamma_{\mu}l_{r})(\overline{d}\gamma^{\mu}d)}}
\newcommand{\QjdOne}{\ensuremath{\mathcal{Q}_{\mathrm{qd}}^{(1)}}}
\newcommand{\QjdOneDef}{\ensuremath{(\overline{q}\gamma_{\mu}q)(\overline{d}\gamma^{\mu}d)}}
\newcommand{\QjuOne}{\ensuremath{\mathcal{Q}_{\mathrm{qu}}^{(1)}}}
\newcommand{\QjuOneDef}{\ensuremath{(\overline{q}\gamma_{\mu}q)(\overline{u}\gamma^{\mu}u)}}
\newlength\cmsTabSkip\setlength{\cmsTabSkip}{1ex}
\newlength\cmsFigWidth
\ifthenelse{\boolean{cms@external}}
{
  \providecommand{\cmsLeft}{upper\xspace}
  \providecommand{\cmsRight}{lower\xspace}
  \setlength{\cmsFigWidth}{0.95\columnwidth}
  \providecommand{\maybeCmsTable}[1]{\resizebox{0.49\textwidth}{!}{#1}}
}
{
  \providecommand{\cmsLeft}{left\xspace}
  \providecommand{\cmsRight}{right\xspace}
  \setlength{\cmsFigWidth}{0.3\textwidth}
  \providecommand{\maybeCmsTable}[1]{{#1}}
}

\cmsNoteHeader{SMP-25-001} 
\title{Search for anomalous couplings in \texorpdfstring{$\PW\PW$}{WW} and \texorpdfstring{$\PW\PZ$}{WZ} production with single-lepton final states in proton-proton collisions at \texorpdfstring{$\sqrt{s} = 13\TeV$}{sqrt(s) = 13 TeV}}
\date{\today}
\abstract{
A search for deviations from the standard model using an effective field theory approach is carried out using the proton-proton collision data set recorded by the CMS experiment at the LHC at a center-of-mass energy of 13\TeV, corresponding to an integrated luminosity of $138\fbinv$.
In this study Wilson coefficients corresponding to dimension-six effective field theory operators that would lead to anomalous gauge boson self-couplings and modified couplings between vector bosons and quarks are constrained. 
Diboson ($\PW\PW$ and $\PW\PZ$) production processes with one $\PW$ boson decaying to a lepton plus antineutrino or charge conjugate and the second $\PW$ or $\PZ$ boson decaying hadronically are considered.
Since the contribution from anomalous couplings is expected to be most visible at high energy scales, the focus is on final states where the hadronic decay products of a $\PW$ or $\PZ$ boson are merged into a single large-radius jet. A dedicated classifier based on machine learning is employed to separate hadronic $\PW$ and $\PZ$ boson decays from background processes.
The most stringent constraints to date on the Wilson coefficients of operators corresponding to anomalous triple gauge boson couplings are reported.
The bounds set on the couplings between vector bosons and quarks are competitive with those from previous inclusive jet measurements.}

\hypersetup{
pdfauthor={CMS Collaboration},
pdftitle={Search for anomalous couplings in WW and WZ production with single-lepton final states in proton-proton collisions at sqrt(s)=13 TeV},
pdfsubject={CMS},
pdfkeywords={CMS, aTGC, SMEFT, boosted vector bosons}}

\maketitle 

\section{Introduction}\label{sec:intro}

The standard model (SM) of particle physics has been remarkably successful at describing the known fundamental particles and their interactions. However, there are several phenomena that it cannot accommodate, such as a description of dark matter. A large number of beyond-the-SM (BSM) models have been proposed, describing BSM physics at energy scales larger than the electroweak (EW) scale.
While it is possible that some predicted heavy BSM resonances could be directly observed, it is also possible that the BSM physics lies at an energy  scale far beyond the direct reach of the CERN LHC experiments. Although the direct production of new heavy BSM particles might not be within the current experimental reach, these particles could have a measurable impact on the interactions of SM particles accessible at the LHC.

In this analysis, we look for such modifications to the production cross sections of $\WW$ and $\WZ$ processes, collectively referred to as $\WV$, through BSM couplings to the gauge bosons.
We probe BSM physics at an energy scale, $\PgL$, that is far larger than the LHC-accessible regime.
We utilize an effective field theory (EFT) approach, where BSM contributions are parameterized in a model-independent fashion as an infinite series of operators suppressed by powers of $1/\PgL$, such that one can recover the SM in the limit $\PgL \to \infty$.
All possible BSM effects are incorporated in an effective Lagrangian ($\mathcal{L}_{\text{EFT}}$),
which can be written as
\begin{equation}
\mathcal{L}_{\text{EFT}} = \mathcal{L}_{\text{SM}} + \sum_{d,i} \frac{c_{i}^{d}}{\Lambda^{d-4}} \mathcal{O}_{i}^{d},
\end{equation}
where $\mathcal{L}_{\text{SM}}$ is the SM Lagrangian
and $\mathcal{O}_{i}^{d}$ are the newly introduced operators of mass dimension $d$.
For each $\mathcal{O}_{i}^{d}$, there exists a dimensionless Wilson coefficient~(WC), $c_{i}^{d}$,
scaling the strength of the respective operator. The SM is defined at $d=4$, and the contributions from higher-dimensional operators are suppressed by $1/\Lambda^{d-4}$.
Neglecting operators violating lepton or baryon number conservation~\cite{HELSET2020135132}, the leading contributions from BSM particles are
expected from operators of mass dimension six~\cite{Buchmuller:1985jz}.

We consider two EFT conventions to parameterize the effect of dimension-six operators on $\WV$ productions.
In the first convention, referred to as the HISZ convention or the \textsc{EFTDim6} model~\cite{Degrande_2013,paper:eft}, we consider five operators that modify the self-couplings of $\PW$ and $\PZ$ bosons, in particular the $\PW\PW\PZ/\PGg$ vertex appearing in $\WV$ production processes. 

These include three operators that conserve charge-parity ($CP$) symmetry,
\begin{equation}
\begin{aligned}
\mathcal{O}_{\PW\PW\PW}^\mathrm{HISZ}
&= \mathrm{Tr}\!\left[\PW_{\mu\nu}\PW^{\nu\rho}\PW^{\mu}{}_{\rho}\right], \\
\mathcal{O}_{\PW}^\mathrm{HISZ}
&= \left(D_{\mu}\Phi\right)^{\dagger}\PW^{\mu\nu}\left(D_{\nu}\Phi\right), \\
\mathcal{O}_{\PB}^\mathrm{HISZ}
&= \left(D_{\mu}\Phi\right)^{\dagger}\PB^{\mu\nu}\left(D_{\nu}\Phi\right).
\end{aligned}
\end{equation}
where $\Phi$ is the Higgs doublet field, $D_{\mu}$ the SM EW covariant derivative, and $\PW_{\mu\nu}$ and $\PB_{\mu\nu}$ are the field-strength tensors of the $\mathrm{SU}(2)_L$  and $\mathrm{U}(1)_Y$ gauge groups, respectively.

This choice of operators can be constrained by only measuring the triple gauge boson couplings~\cite{Grojean:2006nn}.
The three operators have WCs $c_{\PW\PW\PW}^\mathrm{HISZ}$, $c_{\PW}^\mathrm{HISZ}$, and $c_{\PB}^\mathrm{HISZ}$, respectively. 

The two $CP$-violating operators, $\mathcal{O}_{\widetilde{\PW}\PW\PW}^\mathrm{HISZ}$ and $\mathcal{O}_{\widetilde{\PW}}^\mathrm{HISZ}$, are defined as:
\begin{equation}
\begin{aligned}
\mathcal{O}_{\widetilde{\PW}\PW\PW}^\mathrm{HISZ}
&=
\mathrm{Tr}\!\left[
\widetilde{\PW}_{\mu\nu}
\PW^{\nu\rho}
\PW^{\mu}{}_{\rho}
\right], \\
\mathcal{O}_{\widetilde{\PW}}^\mathrm{HISZ}
&=
\left(D_{\mu}\Phi\right)^{\dagger}
\widetilde{\PW}^{\mu\nu}
\left(D_{\nu}\Phi\right).
\end{aligned}
\end{equation}
where  $\widetilde{\PW}^{\mu\nu} = \frac{1}{2}\varepsilon^{\mu\nu\rho\sigma}{\PW}_{\rho\sigma}$ and $\varepsilon$ is the Levi-Civita symbol.

The WCs corresponding to these operators are $c_{{\widetilde\PW}\PW\PW}^\mathrm{HISZ}$ and $c_{\widetilde\PW}^\mathrm{HISZ}$, respectively. These additional contributions to the EW vector boson self-interactions beyond those present in the SM are referred to as anomalous triple gauge couplings (aTGCs).

The second convention is the SM EFT (SMEFT) framework for describing BSM physics effects in tails of the distributions~\cite{Brivio:2020onw,Brivio:2017vri,Passarino:2019yjx,David:2020pzt,daSilvaAlmeida:2018iqo}. In this analysis, we probe 18 SMEFT operators that modify the shape of the invariant mass of the diboson system.
These include operators that  modify the self-couplings of $\PW$ and $\PZ$ bosons and the couplings of
$\PW$ and $\PZ$ bosons to the light-quark generations. In addition, four-fermion operators with leptons and quarks introduce contact interactions between them, which also include the operators with contact interactions between light and third generation quarks.
The probed SMEFT operators are defined in the Warsaw basis~\cite{Grzadkowski:2010es} and are listed in Table~\ref{tab:aC_ops}. We assume the \textsc{topU3L}~\cite{Brivio:2020onw} flavor structure under which
\begin{itemize}
\item{the flavor symmetry is assumed for the first two generations of quarks while quarks of the third generation are described by independent fields,}
\item{right-handed charged currents of the first two generations of quarks are forbidden,}
\item{mixing effects in the quark sector are neglected,}
\item{the three lepton generations are treated symmetrically.}
\end{itemize}
This symmetry restricts the flavor structure of the dimension-six operators and the number of independent WCs is significantly reduced.

\begin{table}[!htb]
\centering
\topcaption{\label{tab:aC_ops}
The dimension-six SMEFT operators $(\mathcal{Q})$ and their corresponding Wilson coefficients ($c$) studied in this analysis,
following the definitions of Refs.~\cite{Brivio:2017btx,Brivio:2020onw}, where $(q\ \mathrm{or}\ u,d)$ denote quark fields of the first two
generations and $(l\ \mathrm{or}\ e,\nu)$ lepton fields of all three generations. The Higgs doublet field is indicated by $H$; $D$ represents a covariant derivative; $\protect\Dbid$
denotes the Hermitian bidirectional covariant derivative;  $X = G, W, B$ denotes a vector boson field strength tensor; $p$ and $r$ are flavor
indices; and $\sigma^{i}$ ($i = 1,2,3$) are the Pauli matrices. Fermion fields are represented by $\psi$, with $L$ and $R$
indicating left- and right-handed fermion fields, respectively.}
\begin{tabular}{cc}
Operator &  Corresponding WC\\ [\cmsTabSkip]
 \hline
\noalign{\vskip 0.5mm} 
\multicolumn{2}{c}{$\boldsymbol{X^{3}}$} \\
\Qw             $= \QwDef$              & \cw        \\ 
\Qwtil          $= \QwtilDef$           & \cwtil     \\
\multicolumn{2}{c}{$\boldsymbol{X^{2} H^{2}}$} \\       
\Qhwb           $= \QhwbDef$            & \chwb      \\
\Qhwbtil        $= \QhwbtilDef$         & \chwbtil    \\
\multicolumn{2}{c}{$\boldsymbol{\psi^{2} H^{2} D}$} \\
\Qhd            $= \QhdDef$             & \chd        \\
\Qhu            $= \QhuDef$             & \chu        \\
\QhjOne         $= \QhjOneDef$          & \chj{1}     \\
\QhjThree       $= \QhjThreeDef$        & \chj{3}     \\
\multicolumn{2}{c}{$\boldsymbol{\psi^{4}}\text{\textbf{, }}\boldsymbol{(\overline{L}L)(\overline{L}L)}$} \\ 
\QjjOneOne      $= \QjjOneOneDef$       & \cjj{1}{1}  \\
\QjdOne         $= \QjdOneDef$          & \cjd{1}     \\
\QljOne         $= \QljOneDef$          & \clj{1}     \\
\QljThree       $= \QljThreeDef$        & \clj{3}     \\
\QjjOneEight    $= \QjjOneEightDef$     & \cjj{1}{8}  \\
\QjjThreeOne    $= \QjjThreeOneDef$     & \cjj{3}{1}  \\ 
\QjjThreeEight  $= \QjjThreeEightDef$   & \cqj{3}{8}  \\
\multicolumn{2}{c}{$\boldsymbol{\psi^{4}}{\textbf{, }}\boldsymbol{(\overline{L}L)(\overline{R}R)}$} \\        
\Qlu            $= \QluDef$             & \clu        \\
\Qld            $= \QldDef$             & \cld        \\
\QjuOne         $= \QjuOneDef$          & \cju{1}     \\ [\cmsTabSkip]
\end{tabular}
\end{table}

\begin{figure}[htb]
\centering
\includegraphics[width=0.49\textwidth]{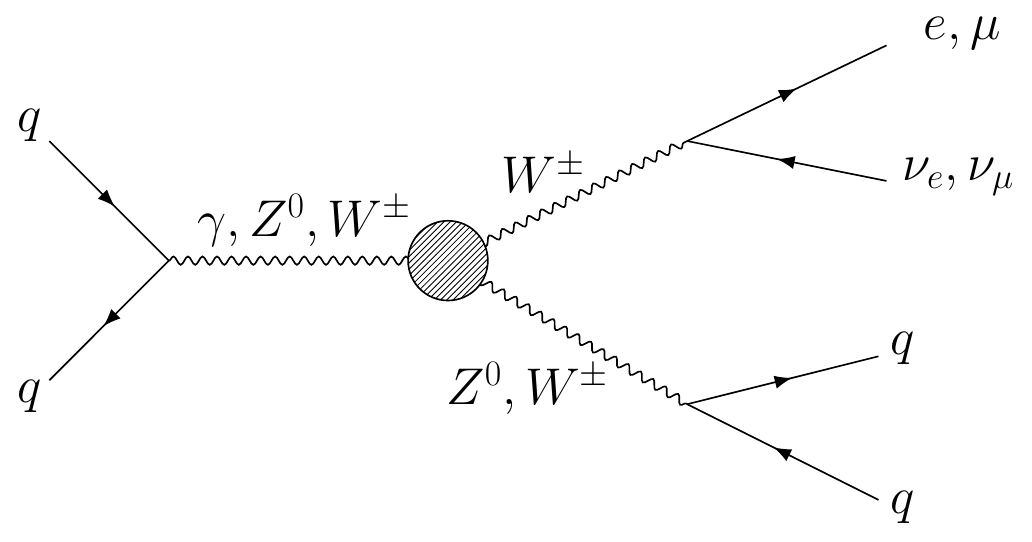}
\caption{The leading order Feynman diagram for the process studied in this analysis, with an anomalous triple gauge coupling vertex, where one $\PW$ boson decays to a lepton and a neutrino and another $\PW$ ($\PZ$) boson decays to quarks.}\label{fig:FeynmanDiagram}
\end{figure}

The leading order (LO) Feynman diagram for the process under study, where aTGCs may contribute, is shown in Fig.~\ref{fig:FeynmanDiagram}.
We consider the single-charged-lepton final state, where one $\PW$ boson decays into a charged lepton (an electron or a muon, including the contributions from leptonic $\tau$ decays) and a neutrino, and the other vector boson ($\PW$ or $\PZ$) decays into quarks.

In comparison with a fully leptonic final state, the branching fraction is increased, whilst cases with only one neutrino allow for a full reconstruction of the diboson system.
Unlike the fully hadronic final state, the presence of a charged lepton significantly
reduces the contribution of background events composed of jets produced via strong interactions, \ie, the quantum chromodynamics (QCD) multijet background.
The effect of aTGCs is strongest at larger boson
momenta and diboson invariant masses. Thus, we consider the scenario where the hadronically decaying boson has a significant Lorentz boost, and the two jets from the gauge boson decay will overlap
in the detector to form a single large-radius jet. To improve sensitivity to gauge boson decays and to reject other background processes, jet substructure techniques are used.
We analyze spectra of the reconstructed boson pair invariant mass $\mwv$ to look for deviations from the SM expectation.
As the jet mass resolution is limited, jets from $\PW$ and $\PZ$ bosons cannot be uniquely differentiated, and thus contributions from the various aTGCs must be considered simultaneously.

Previous searches for such signatures with the ATLAS and CMS experiments have been performed in leptonic~\cite{Chatrchyan:2012sga,Chatrchyan:2013yaa,CMS:2014xja,Khachatryan:2015sga,Khachatryan:2016poo,Sirunyan:2017zjc,Aad:2011cx,Aad:2011xj,Aad:2012rxl,Aad:2012twa,ATLAS:2012mec,Aad:2012awa,Aad:2016wpd,Aad:2016ett,ATLAS:2012mec,Aad:2016wpd,Aad:2016ett,Aaboud:2016urj,Aaboud:2017rwm,Sirunyan:2019bez} and single-charged-lepton final states~\cite{Chatrchyan:2012bd,Sirunyan:2017bey,Aad:2014mda,Aaboud:2017cgf,CMS:2019ppl}.
This work builds upon a previous aTGC analysis using data corresponding to an integrated luminosity of 36\fbinv collected by the CMS experiment during proton-proton ($\Pp\Pp$) collisions at $\sqrt{s} = 13\TeV$ in 2016~\cite{CMS:2019ppl}, which provided the most stringent limits at that time.
The analysis uses the full data set recorded in 2016--2018 by the CMS experiment, corresponding to 138\fbinv, and employs improved jet substructure techniques and background estimation procedures. Significant improvement in the analysis is also achieved by the usage of a dynamic graph convolutional neural network (\textsc{ParticleNet}~\cite{paper:PN}) for the identification of $\PW$ and $\PZ$ boson jets.
Tabulated results are provided in the HEPData record for this work~\cite{HEPData}.

\section{The CMS detector and simulated samples}\label{sec:MC}

The central feature of the CMS apparatus is a superconducting solenoid of 6\unit{m} internal diameter, providing a magnetic field of 3.8\unit{T}. A silicon pixel and strip tracker, a lead tungstate crystal electromagnetic calorimeter (ECAL) with a silicon strip preshower detector (ES) placed at the front faces of the two endcap calorimeters, and a brass and scintillator hadron calorimeter (HCAL), each composed of a barrel and two endcap sections, are installed inside the solenoid. Forward calorimeters extend the pseudorapidity ($\eta$) coverage provided by the barrel and endcap detectors. Muons are detected in gas-ionization chambers embedded in the steel flux-return yoke outside the solenoid. 
A more detailed description of the CMS detector, together with a definition of the coordinate system and the relevant kinematic variables, can be found in Ref.~\cite{CMS:2008xjf,CMS:2023gfb}.

Events of interest are selected using a two-tiered trigger system~\cite{CMS:2016ngn}. The first level, composed of custom hardware processors, uses information from the calorimeters and muon detectors to select events at a few 100\unit{kHz} within a fixed latency of about 4\mus~\cite{CMS:2020cmk}. The second level, known as the high-level trigger, consists of a farm of processors running a version of the full event reconstruction software optimized for fast processing, and reduces the event rate to a few \unit{kHz} before data storage~\cite{CMS:2016ngn,CMS:2024aqx}.

Signal and background processes are simulated using multiple Monte Carlo (MC) event generators and are reconstructed using the same procedures as employed for data.
The simulated samples are used to optimize the event selection, evaluate selection efficiencies and systematic uncertainties, and compute expected event yields.

The SM \WV samples are produced with the \MGvATNLO~(v2.6.5)~\cite{MadGraph5_aMCatNLO} generator at next-to-LO (NLO) in QCD.
The production of $\ttbar$ pairs, single top quarks in the $t$-channel, and $\PQt\PW$ pairs uses the \POWHEG (v2) event generator~\cite{POWHEG2,POWHEG3}.
The \MGvATNLO generator is also used to simulate single top quark production in the $s$-channel, as well as $\PW$ boson production in association with jets ($\Wj$) at NLO in QCD with up to two additional partons in the matrix element calculations. Other SM background processes simulated using \POWHEG include associated production of a $\PV$ boson with a Higgs boson ($\PHiggs$), are referred to as ``other'' henceforth.

Two sets of \WV signal samples are generated using \MGvATNLO using two different EFT simulation packages.
The first is generated at NLO accuracy in QCD with up to one additional parton at the matrix-element level using the \textsc{EFTDim6} model.
The second is generated at LO with up to one additional parton in the matrix element calculations using the \textsc{SMEFTsim} package~\cite{Brivio:2017btx}.
For both signal samples, a reweighting method~\cite{Mattelaer:2016gcx} implemented in \MGvATNLO is used to reproduce the effect of different WC values on the event kinematics.
The total matrix elements are calculated at several points in the EFT parameter space
using the relevant EFT simulation package via

\begin{equation}
\begin{aligned}
\abs{\mathcal{M}}^{2}
={}&
\abs{\mathcal{M}_{\mathrm{SM}}}^{2}
+\sum_{i} 2c_i\,\mathrm{Re}\!\left(
\mathcal{M}_{\mathrm{SM}}^{*}
\mathcal{M}_{\mathrm{BSM}}^{i}
\right)
\\
&+\sum_{i}\sum_{j}
c_i c_j\,\mathrm{Re}\!\left(
\mathcal{M}_{\mathrm{BSM}}^{i*}
\mathcal{M}_{\mathrm{BSM}}^{j}
\right).
\end{aligned}
\end{equation}
where $\mathcal{M}_\text{SM}$ and $\mathcal{M}_\text{BSM}$ are the matrix element contributions
from SM and BSM physics, respectively. The WCs, $c_i$ and $c_j$, are assumed to be real and the sum runs over all considered EFT operators.
Weights corresponding to the ratios $\abs{\mathcal{M}}^{2}/\abs{\mathcal{M}_\text{SM}}^{2}$
for selected base points in the EFT parameter space are stored for each simulated event.
With $(N+1)(N+2)/2$ of these weights,
the polynomial for $N$ EFT operators is solved to extract weights,
containing the full dependence on the WCs, on an event-by-event basis.
In the computation of $\abs{\mathcal{M}}^{2}$, we only consider diagrams with a single EFT vertex, such that terms
linear in $c_i$ encode the SM-BSM interference and quadratic terms describe the pure BSM contribution.
Setting all WCs to zero, the LO samples with an additional parton are required to reproduce the SM
distributions modeled at NLO in perturbative QCD.
This is done by reweighting the LO samples with the NLO-to-LO QCD scale factors that account for the differences observed in the shape of $\mwv$ between the LO and NLO simulations of SM \WV.
In the phase space of interest to this analysis, with $\mwv > 950\GeV$, we find scale factors ranging from 2.8 to 5, rising as a function of $\mwv$.
Given those large correction factors, we do not assume the computed $\text{SM}$ cross section $\sigma_\text{SM}$, but instead estimate it from data.
We assume that higher-order QCD corrections factorize from the EFT effects such that each $\mwv$ bin cross section is
$\sigma_\text{SM} (1 + c_i A_i + c_i c_j B_{ij})$, where the $\sigma_\text{SM}$ is obtained from the fit to data, and $A_i$ and $B_{ij}$ are constants derived from the reweighted EFT simulation.
Signal samples are generated in bins of $\mwv$ to ensure that sufficient events populate the tail of the distribution.
The EFT contributions to processes other than \WV are not considered in this analysis.
This is a good assumption in the HISZ model that considers only triple gauge boson couplings, which do not affect the dominant background processes. It is, however, not necessarily guaranteed that all considered SMEFT operators do not affect the shape of the background processes, thus is only approximate in this case.

Parton showering and hadronization are modeled using the generator \PYTHIA~8.309~\cite{PYTHIA8:2} with the tune \textsc{CP5}~\cite{CP5tune_CMS} for all simulated samples.
The matrix element calculation and parton shower simulation are matched using the
FxFx~\cite{Frederix:2012ps} and MLM~\cite{Alwall:2007fs} algorithms for the processes
simulated with \MGvATNLO at NLO and LO, respectively.
All simulated samples use the NNPDF3.1~\cite{NNPDF:2017mvq} parton distribution functions~(PDFs) at next-to-NLO (NNLO) accuracy.
The generated events are then passed through the \GEANTfour~\cite{GEANT4:2002zbu} package to simulate the detector response before
they are treated with the same reconstruction algorithms as the real data.
In addition, $\Pp\Pp$ inelastic collisions simulated with \PYTHIA are mixed into each event
to model the effect from additional collisions within the same or adjacent bunch crossings~(pileup).
Simulated events are reweighted to match the pileup profile of data.
The simulated events are corrected for various detector effects that are not well modeled in the simulation, which will be given in the following section.

Most of the cross sections used to normalize the MC samples have been calculated at NNLO precision in perturbative QCD.
The NNLO cross section for $\Wj$, corresponding to a NNLO-to-NLO QCD $k$-factor of 1.018, is calculated with \textsc{fewz} (v3.1)~\cite{FEWZ}.
The $\ttbar$ cross section is calculated at NNLO precision with \textsc{Top++} (v2.0)~\cite{xsecTTBar,Czakon:2013goa}.
The $\WW$ and $\WZ$ cross sections are taken from Refs.~\cite{xsecWW} and~\cite{xsecWZ}, both at NNLO precision.
Single top quark cross sections are calculated with the \textsc{Hathor}~(v2.1) program~\cite{xsecSingleTop}.
The NLO \POWHEG{+}\PYTHIA\ $\ttbar$ sample used in the analyses shows deviations in the transverse momentum ($\pt$) spectrum of top quarks when compared with data~\cite{TopPTdilepton,CMS:2016oae}. The $\ttbar$ events are thus reweighted based on the measurements of the top quark $\pt$ spectrum, to account for missing higher-order effects.

\section{Event reconstruction and calibration}\label{eventreco}

A particle-flow (PF) algorithm~\cite{CMS:2017yfk} aims to reconstruct and identify each individual particle in an event, with an optimized combination of information from the various elements of the CMS detector. The reconstructed particles are classified as charged hadrons, neutral hadrons, photons, muons, and electrons, providing an optimized estimate of their momenta and energies. The energy of photons is obtained from the ECAL measurement. The energy of electrons is determined from a combination of the electron momentum at the primary interaction vertex as determined by the tracker, the energy of the corresponding ECAL cluster, and the energy sum of all bremsstrahlung photons spatially compatible with originating from the electron track. The energy of muons is obtained from the curvature of the corresponding track. The energy of charged hadrons is determined from a combination of their momentum measured in the tracker and the matching ECAL and HCAL energy deposits, corrected for the response function of the calorimeters to hadronic showers. Finally, the energy of neutral hadrons is obtained from the corresponding corrected ECAL and HCAL energies. The primary vertex (PV) is taken to be the vertex corresponding to the hardest scattering in the event, evaluated using tracking information alone, as described in Section 9.4.1 of Ref.~\cite{CMS-TDR-15-02}.

For each event, hadronic jets are clustered from PF candidates using the infrared- and collinear-safe anti-\kt algorithm~\cite{Cacciari:2008gp, Cacciari:2011ma} with a certain distance parameter (R).
In this analysis two collections of jets with $R=0.4$ (small-$R$) and $R=0.8$ (large-$R$) are used.
Jet momentum is determined as the vectorial sum of all particle momenta in the jet, and is found from simulation to be, on average, within 5 to 10\% of the true momentum
over the entire $\pt$ spectrum and detector acceptance.
Pileup can contribute additional tracks and calorimetric energy depositions to the jet momentum.
For $R=0.4$ jets this effect is mitigated by discarding charged particles identified to be originating from pileup vertices from jet clustering and applying an offset to correct for remaining contributions.
For $R=0.8$ jets, the more sophisticated pileup-per-particle identification algorithm (PUPPI)~\cite{Sirunyan:2020foa,Bertolini:2014bba} is used to mitigate the effect of pileup at the reconstructed-particle level before jet clustering, making use of local shape information, event pileup properties, and tracking information. A local shape variable is defined, which distinguishes between collinear and soft diffuse distributions of other particles surrounding the particle under consideration. The former is attributed to particles originating from the hard scatter and the latter to particles originating from pileup interactions. Charged particles identified to be originating from pileup vertices are discarded. For each neutral particle, a local shape variable is computed using the surrounding charged particles compatible with the PV within the tracker acceptance ($\abs{\eta} < 2.5$), and using both charged and neutral particles in the region outside of the tracker coverage. The momenta of the neutral particles are then rescaled according to their probability to originate from the PV, deduced from the local shape variable, superseding the need for jet-based pileup corrections~\cite{Sirunyan:2020foa}. 
Finally, jet energy corrections are derived from simulation to bring the measured response of jets to that of particle-level jets on average.
In situ measurements of the momentum balance in dijet, photon+jet, $\PZ$+jet, and multijet events are used to account for any residual differences in the jet energy scale (JES) between data and simulation~\cite{CMS:2016lmd}. The jet energy resolution (JES) amounts typically to 15--20\% at $30\GeV$, 10\% at $100\GeV$, and 5\% at $1\TeV$~\cite{CMS:2016lmd}.

Additional selection criteria are applied to remove jets (both small-$R$ and large-$R$) potentially dominated by anomalous contributions 
from various subdetector components or reconstruction failures~\cite{Sirunyan:2020foa}.
In 2018, two sectors of the HCAL endcap in the region with azimuthal angle $-1.57< \phi< -0.87$ and $-3.2< \eta< -1.3$ were not functional 
for the last ${\approx}$65\% of the data taking. In data recorded during this time, events are rejected if 
the leading jet is reconstructed within this region. Events in the simulation are weighted accordingly if the 
leading jet is reconstructed in this region to reflect the resulting inefficiency.

To reject events involving top quarks, the small-$R$ jets with $\pt>30\GeV$ and $\abs{\eta}<2.5$ originating from $\PQb$ quarks are identified ($\PQb$-tagged) by a multivariate algorithm called \DeepJet~\cite{BTV-16-002,CMS:deepjet}. For the chosen working point, the efficiency to select $\PQb$\ jets is about 70--80\% and the rate for incorrectly tagging jets originating from the hadronization of gluons or $\cPqu$, $\cPqd$, $\cPqs$ quarks is $\sim$1\%. The efficiency and mistag rate of $\PQb$ tagging in simulation is corrected to match the data.

The large-$R$ jet with largest (leading) $\pt$ in the event is reclustered with the Cambridge--Aachen (CA) algorithm~\cite{Dokshitzer:1997in}.
Then, the modified mass-drop algorithm~\cite{Dasgupta:2013ihk,Butterworth:2008iy}, known as the soft-drop algorithm~\cite{Larkoski:2014wba}, is applied to compute the soft-drop jet mass $\msd$.
This splits the jet into two subjets by undoing the last step of the CA jet clustering.
It regards the jet as the final soft-drop jet if the two subjets satisfy the condition:
\begin{equation}
 \frac{ \min ( \pt^{(1)}, \pt^{(2)} ) }{\pt^{(1)} + \pt^{(2)}  } > z_{\text{cut}} \Big(  \frac{ \Delta R_{12} }{R} \Big) ^ {\beta_{\text{sd}}},
\end{equation}
where $\pt^{(1)}$ and $\pt^{(2)}$ are the transverse momenta of the two subjets, $R$ is the jet radius parameter, $\Delta R_{12} = \sqrt{\smash[b]{(\Delta y_{12})^2 + (\Delta \phi_{12})^2}}$ is the distance between the two subjets, and $z_{\text{cut}} $ and $\beta_{\text{sd}}$ are tuneable parameters of the soft-drop algorithm, set to default parameters $z_{\text{cut}} = 0.1$ and $\beta_{\text{sd}} = 0$ used in CMS studies~\cite{CMS:2020poo}.
If the condition is not met, the subjet with the lower $\pt$ is rejected.
The declustering procedure is repeated splitting the higher $\pt$ subjet into two subjets by undoing another CA clustering step, iterating until the condition is met.
This soft-drop procedure removes soft and wide-angle radiation from the jet, thereby making the jet mass observables more resilient to effects from pileup, the underlying event, and initial-state radiation.

The \textsc{ParticleNet}~\cite{paper:PN} tagger is used to discriminate between large-$R$ jets originating from a heavy vector boson $\PV$ (\PW or $\PZ$) and those from QCD multijet production.
This tagger exploits the jet substructure variables for the reconstruction and identification of the $\PV \to \PQq\PQq $ candidates. This procedure is referred to as $\PV$ boson tagging hereafter.
At its core, the \textsc{ParticleNet} algorithm treats a jet as an unordered set of its constituents and builds a permutation-invariant graph neural network to effectively exploit the correlation between the jet constituents. Jet constituent PF candidates and secondary vertices within the jet cone are used as inputs to the algorithm. A mass-decorrelated version of \textsc{ParticleNet} is used (\textsc{ParticleNet}-MD), which is trained using a signal sample containing a generic particle $X$ with a flat mass spectrum, which decays into two highly Lorentz-boosted particles.
The \textsc{ParticleNet}-MD algorithm is a multiclass tagger that provides different output nodes: 
$X \to \PQc\PQc$ (``Xcc"), $X \to \PQq\PQq$ (``Xqq"), and $X \to \text{QCD}$ (``QCD"), where $\PQq$ are light-flavor jets. To separate signal from QCD-induced jets, they are combined in the following way:
\begin{equation}
\PW/\PZ \text{ tagging score} = \frac{\mathrm{Xcc} + \mathrm{Xqq}}{\mathrm{Xcc} + \mathrm{Xqq} + \mathrm{QCD}}.
\end{equation}

This $\PV$ boson candidate jet must have $75<\msd<115\GeV$ and should pass the ``medium'' working point of the \textsc{ParticleNet} tagger.
The selected $\msd$ range encloses the mean reconstructed $\PW$ and $\PZ$ masses, which are about 10\GeV larger than the true $\PW$ and $\PZ$ masses~\cite{ParticleDataGroup:2024cfk} due to residual effects from pileup, underlying event, and initial-state radiation, remaining after the soft-drop and PUPPI algorithms.
The chosen ``medium'' working point of the \textsc{ParticleNet} tagger corresponds to a $\PV$ boson tagging efficiency of 43\% and probability of misidentifying a light-flavor quark or gluon jet as a $\PV$ boson candidate jet of 1\%, as determined in a sample of simulated $\ttbar$ events.

Electrons and muons are reconstructed by associating a track reconstructed in the tracking detectors with either a cluster of energy in the ECAL~\cite{Khachatryan:2015hwa,CMS-DP-2020-021} or a track in the muon system~\cite{Sirunyan:2018fpa}. Electron and muon candidates must pass certain identification criteria to be further selected in the analysis. A ``loose'' identification is defined by requiring $\pt > 10\GeV$ and $\abs{\eta} < 2.5$ (2.4) for electrons (muons). Furthermore, electrons are required not to have $1.44 < \abs{\eta} < 1.57$ that corresponds to the ECAL transition region.
A ``tight'' selection is then required following the definitions provided in Refs.~\cite{Khachatryan:2015hwa,Sirunyan:2018fpa}, including requirements on the impact parameter of the candidates with respect to the PV and their isolation with respect to other particles in the event~\cite{Sirunyan:2018egh}. 

The missing transverse momentum vector \ptvecmiss is computed as the negative vector sum of the transverse momenta of all the PF candidates in an event, and its magnitude is denoted as $\ptmiss$~\cite{CMS:2019ctu}. The \ptvecmiss is modified to account for corrections to the energy scale of the reconstructed jets in the event. The PUPPI algorithm is applied to reduce the pileup dependence of the \ptvecmiss observable. The \ptvecmiss is computed from the PF candidates weighted by their probability to originate from the PV~\cite{CMS:2019ctu}.  Anomalous high-$\ptmiss$ events can be due to a variety of reconstruction failures, detector malfunctions or noncollision backgrounds. Such events are rejected by event filters that are designed to identify more than 85--90\% of the spurious high-$\ptmiss$ events with a mistagging rate less than 0.1\%~\cite{CMS:2019ctu}. 

\section{Event selection}\label{sec:evtsel}

The data were collected by the CMS experiment in the years 2016 to 2018 and correspond to an integrated luminosity of 138\fbinv.
The analysis employs triggers requiring an isolated electron (muon) in the high-level trigger, with an online $\pt$ threshold ranging between 27--35 (24--27)\GeV.
This ensures a trigger efficiency above 95\% for events that satisfy the offline selection.

This analysis aims to reconstruct single-charged-lepton decays of $\PW\PW$ and $\PW\PZ$ processes, where both vector bosons are boosted, resulting in one isolated charged lepton, $\ptmiss$, and one large-$R$ jet.
A set of selection requirements are applied to either isolate the signal topology while reducing the background processes, or to select background-enriched control regions (CRs) to constrain their normalizations. Background CRs are designed to have minimal signal contamination.
Events are categorized based on the lepton flavor. The analysis strategy is to simultaneously fit the $\mwv$ distribution in all these regions to evaluate signal contributions and to control the normalizations of the background components.

Selected events must contain exactly one lepton (electron or muon) with $\pt>50\GeV$. Events are required to have $\ptmiss>100\GeV$ to reduce the contribution from the QCD multijet background.
Events containing a loosely identified second lepton with $\pt>10\GeV$ are rejected.

A leptonically decaying $\PW$ boson ($\PW^{\Pell}$) is reconstructed using its decay products, \ie, a lepton in combination with \ptmiss.
To reconstruct $\PW^{\Pell}$, we assume that the reconstructed $\ptmiss$ in the event is entirely due to the transverse component of the neutrino momentum.
One can then reconstruct the longitudinal neutrino momentum, $p_z^\PGn$, by using the known $\PW$ boson mass as the invariant mass of the lepton and neutrino four-vectors.
In the case of a complex solution, only the real part is assigned as the longitudinal momentum.
In the case of two real solutions for $p_z^\PGn$, the 
solution minimizing $\abs{p_z^\PGn - p_z^\Pell}$ is chosen. This criterion exploits 
the fact that the charged lepton and neutrino arise from the decay of the same
$\PW$ boson and thus share the same Lorentz boost along the beam axis. In the simulated SM $\WV$ samples, this method assigns the correct solution in $\approx$90\% of the events. The inefficiency partly stems from the events with $\PW \to \PGt\PGnGt \to \Pe\PGne/\PGm\PGnGm+\PGnGt$ decays which are not efficiently reconstructed because of the presence of an additional neutrino. We require the reconstructed $\Wlep$ to have $\pt > 200\GeV$. Events are required to have at least one $\PV$ boson candidate jet with $\pt > 200\GeV$, and the highest \pt $\PV$ jet is selected.

To target the back-to-back topology of diboson events, the following requirements are placed on the angular separations between the reconstructed objects:
\begin{itemize}
\item $\DR(\Pell, \PV) > \pi/2$,
\item $\abs{\Delta\phi(\PV, \ptmiss)} > 2$,
\item $\abs{\Delta\phi(\PV, \Wlep)} > 2$.
\end{itemize}
Furthermore, the invariant mass of the $\WV$ system, $\mwv$, must satisfy the requirement $950 < \mwv < 4550\GeV$.
The lower threshold of $950\GeV$ is chosen such that we have a smoothly falling $\mwv$ spectrum after all selection criteria. The upper limit of $4550\GeV$ is determined by the observed number of data events in the CRs necessary to constrain the backgrounds. With this selection we avoid a turn-on behavior for background processes resulting from the combined efficiency of kinematic, topological and tagger requirements.

To reduce top quark background contributions, events with one or more $\PQb$-tagged small-$R$ jets are rejected.
This set of criteria defines the signal region (SR) event selection. To better distinguish the $\WW$ and $\WZ$ final states, the SR is subdivided into the $\WW$-sensitive region 
($75<\msd<95\GeV$) and the $\WZ$-sensitive region ($95<\msd<115\GeV$), referred to as low and high SRs, respectively.

The $\Wj$ CRs have the same selection criteria as the SR, but with modified $\msd$ requirements, referred to as sidebands. The large-$R$ jet must have $45 <\msd < 75\GeV$ ($115 < \msd < 250\GeV$)  for the low (high) $\Wj$ sideband CR. These regions are studied before and after application of the \textsc{ParticleNet} selection on the $\PV$ boson candidate.
The top quark CR has the same selection as the SR, but inverting the $\PQb$ tag requirement, \ie, the events must have at least one $\PQb$-tagged small-$R$ jet. The $\msd$ requirement is also enlarged to $45 < \msd < 250\GeV$ for the top quark CR.

\section{Background estimation and systematic uncertainties}

A combination of methods based on simulation and CRs in data are used to estimate background contributions, as described in Section~\ref{sec:evtsel}. The dominant background contributions in the SR stem from $\Wj$ followed by $\ttbar$ processes. The $\Wj$ process contributes to the SR when either a quark- or a gluon-initiated large-$R$ jet is mistagged as a $\PV$ boson candidate.
The mistagging rate of a neural-network-based jet substructure algorithm is difficult to predict in simulation, as it also relies on the nonperturbative models of the jet formation~\cite{CMS:2020poo}. Therefore, we estimate the $\Wj$ contribution from data, by constraining the normalization of the process using the $\Wj$ CRs (with a requirement of the tagged large-$R$ jet) during the simultaneous maximum likelihood (ML) fit with the SR. The $\Wj$ low and high sideband CRs are defined, around the SR $\msd$ window, to minimize extrapolation bias from the CR to the SR.
By using the mass-decorrelated tagger \textsc{ParticleNet}-MD, we ensure that the $\msd$ for the $\Wj$ process remains smoothly falling after tagging the large-$R$ jet, enabling this interpolation between the CRs to the SR.
The normalization of $\Wj$ is allowed to vary freely during the fit, to estimate the mistag rate of large-$R$ quark or gluon jets simultaneously with the signal extraction.

The production of $\ttbar$ can yield a leptonic $\PW$ boson and a large-$R$ $\PW$ boson jet, similar to the diboson processes.
The MC simulated samples provide a fair description of the $\ttbar$ process. 
Nevertheless, in order to further constrain the uncertainties related to the modeling of the top quark \pt spectrum and cross section, we also use the $\ttbar$ CR in the final fit to data.
The remaining backgrounds are estimated using dedicated simulated samples, as mentioned in Section~\ref{sec:MC}. This includes the SM production of $\WW$ and $\WZ$, single top quark production ($9\%$), and the background production processes labeled as ``other''. The cross section for SM diboson ($\WW$ and $\WZ$) production is considered a free parameter and estimated in the simultaneous fit of signal and backgrounds to the data in the SRs and CRs. The shape is taken from MC simulation.
The analysis is thus insensitive to a mismodeling of the SM diboson production cross section, but sensitive to BSM contributions that modify the shape of the diboson invariant mass spectrum.

Multiple sources of systematic uncertainties are considered in the analysis. The uncertainties are taken into account through nuisance parameters in the profile likelihood fit described in Section~\ref{sec:results}. These can affect either the rate or both the rate and shape of the $\mwv$ distribution of the signal or background processes.
For each source, the simulated event samples are used to construct corresponding template histograms that describe the expected signal and background distributions for a specific nuisance parameter variation. In the fit of the templates to the data, the best values for both the parameter of interest (\ie, the WC) and all nuisance parameters are determined simultaneously. Independent sources of uncertainties are treated as uncorrelated.

The integrated luminosities for the 2016, 2017, and 2018 data-taking years have 0.82--1.2\% individual uncertainties~\cite{lumi16,CMS:LUM-20-001}, whereas the overall uncertainty for the 2016--2018 period is 0.73\%. The uncertainty in the simulation of pileup events is evaluated by varying the total inelastic pp cross section of 69.2\unit{mb} by $\pm$5\%~\cite{CMS:2018mlc}, which has an impact in the expected signal and background yields of about 1\%. Discrepancies in the lepton reconstruction and identification efficiencies between data and simulation are corrected by applying scale factors to all simulation samples. These scale factors, which depend on the $\pt$ and $\eta$ for both electrons and muons, are determined using the tag-and-probe technique in $\PZ/\PGg \to \Pell\Pell$ events~\cite{Sirunyan:2018fpa,Khachatryan:2015hwa,CMS:2019raw}, with associated uncertainties. The uncertainty in the determination of the trigger efficiency leads to an uncertainty $\approx$0.5\% in the expected signal yield.  In 2016 and 2017, a fraction of trigger primitives in the ECAL was associated with the wrong bunch crossing, leading to a trigger mistiming and a nonnegligible decrease in the trigger efficiency that is not modeled in the simulated samples~\cite{CMS:2024ppo}. Events have been corrected for the mistiming efficiency loss with a per-event weight, and the corresponding uncertainties have been propagated throughout the analysis chain. They are considered uncorrelated among the two data-taking years. 

The JES uncertainties are evaluated by shifting the correction applied to the \pt of the reconstructed jets by one standard deviation~\cite{CMS:2016lmd}, corresponding to multiple independent sources of uncertainties which are partially correlated among different data sets.  The JER systematic uncertainties are evaluated by smearing the four-momenta of the reconstructed jets within uncertainties. This uncertainty is considered uncorrelated for all data-taking periods. Both JES and JER uncertainties are propagated to \ptmiss and the $\msd$.
Additionally, uncertainties in the \ptmiss are calculated by varying the momenta of unclustered particles that are identified neither 
as a jet nor as a lepton.
The $\PV$ boson tagging variables are calibrated in a top quark-antiquark sample enriched in hadronically decaying $\PW$ bosons~\cite{CMS:2020poo}.
The jet $\pt$- and $\eta$-dependent $\PV$ boson tagging corrections with their corresponding uncertainties are propagated to the $\mwv$ distribution. Uncertainties in the $\PQb$ tagging and mistagging data-to-simulation scale factors are applied to reproduce the corresponding efficiencies measured in the data, and are implemented in the ML fit as correlated among the data-taking years.

The dominant theoretical uncertainty arises from the choice of renormalization and factorization scales in the MC event simulation. The theoretical uncertainties are estimated by varying these scales independently up and down by a factor of two from their nominal values  (excluding the two extreme variations) and taking the largest cross section variations as the uncertainty~\cite{Kalogeropoulos:2018cke}. For the signal process and SM $\WW$ and $\WZ$ backgrounds, only shape effects are included, since their normalization is directly measured from observed events in the fit. Both the shape and normalization effects are included for other backgrounds. This uncertainty is considered uncorrelated among the different processes and correlated among the data-taking years. The uncertainty in the modeling of the parton shower is also included, by using the weights corresponding to variations of the strong coupling \alpS in the description of initial- and final-state radiation in the parton showering programs, and is treated as uncorrelated among various processes. The PDF-related uncertainties are evaluated from the eigenvalues of the PDF set following the PDF4LHC prescription~\cite{Butterworth:2015oua}, accounting for the standard deviation estimated from MC replicas.
The uncertainty is treated as uncorrelated between different processes, but correlated across years. Finally, a systematic uncertainty is applied to the top quark \pt reweighting~\cite{Czakon:2017wor,CMS:2016oae}. For the signal and background processes, the limited MC sample size limits the precision of the modeling. The corresponding statistical uncertainties are estimated using the Barlow--Beeston ``lite'' method~\cite{Barlow:1993dm,Conway:2011in}, and are therefore taken as systematic uncertainties applied to each bin of the corresponding distribution.

\begin{figure*}[!htb]
  \centering
\includegraphics[width=0.32\textwidth]{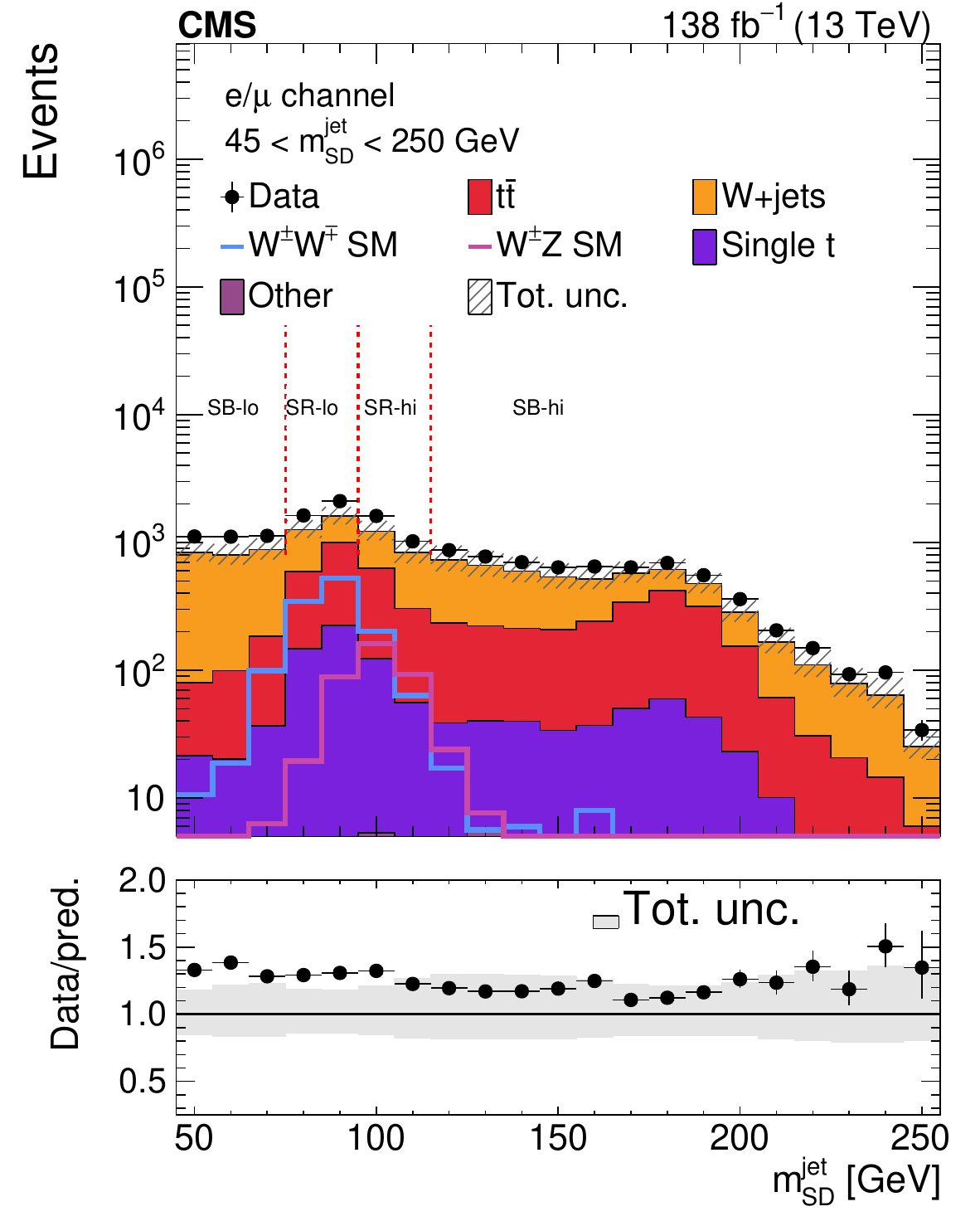}
\includegraphics[width=0.32\textwidth]{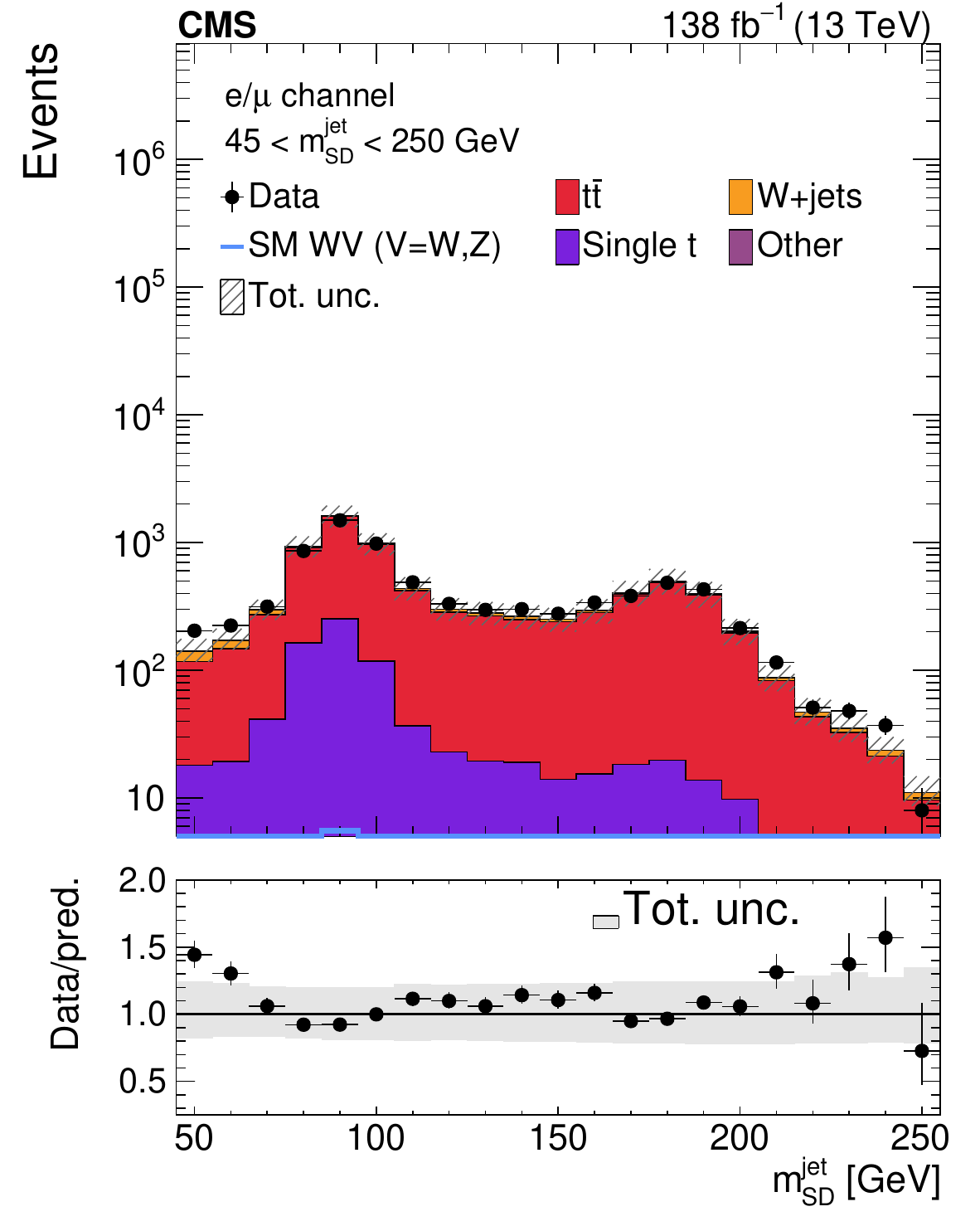}
\includegraphics[width=0.32\textwidth]{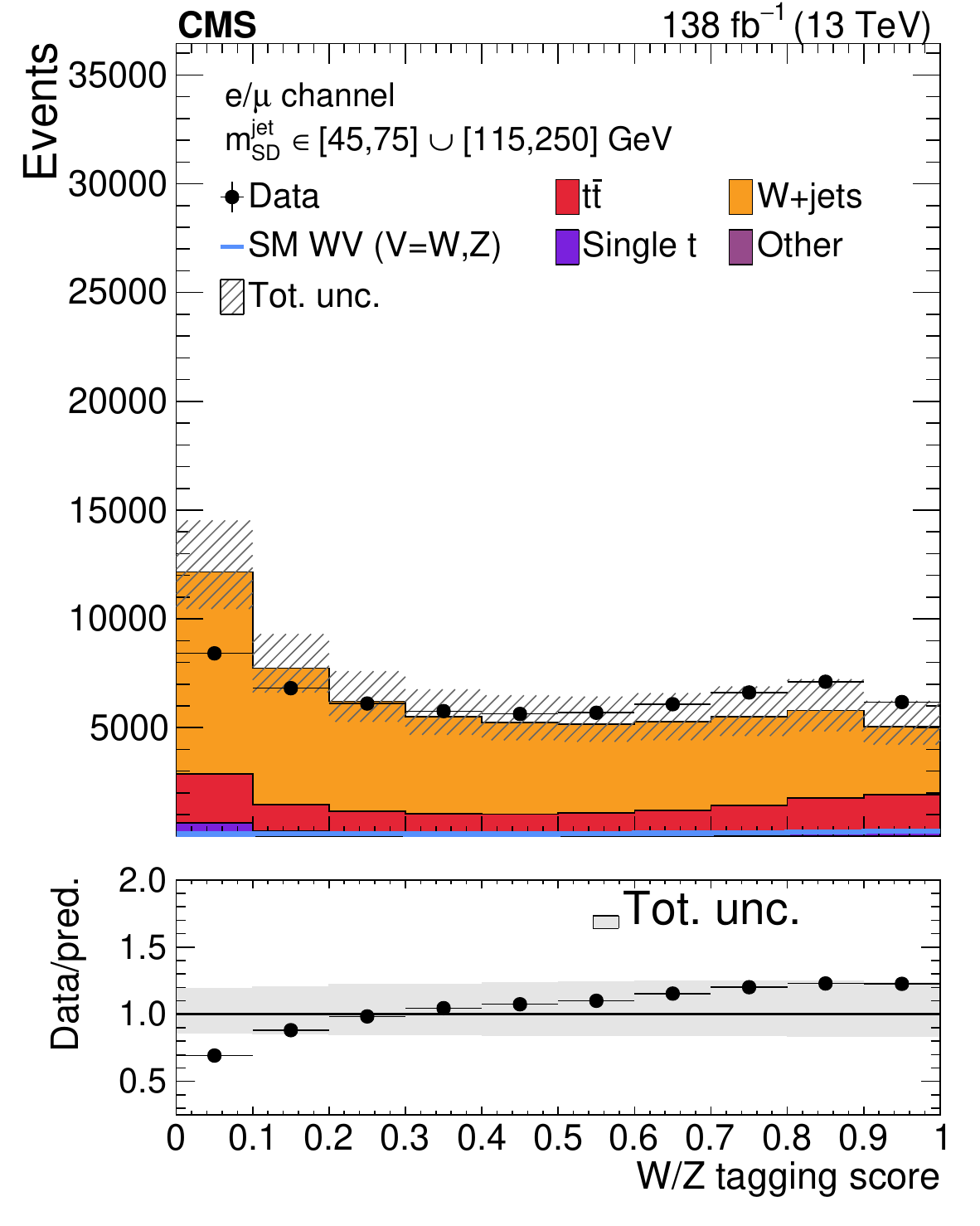}
\caption{Large-$R$ jet $\msd$ in the combined $\Wj$ CR and the signal region (left) and in the $\ttbar$ control region (center) after applying the tagger. The $\PW/\PZ$ boson tagging score in the $\Wj$ CR without tagger requirement is shown in the right panel. These pre-fit distributions are obtained after combining the electron and muon channels and, the data-taking years. The vertical error bars on the data points represent the statistical uncertainty of the data. The uncertainty band shown in the ratio plot contains both statistical and systematic components on the predictions.}\label{fig:wj_FR2_el_nT}
\end{figure*}

Figure~\ref{fig:wj_FR2_el_nT} shows the $\msd$ distribution, after applying a requirement on the tagger, for the large-$R$ jet in the combined $\Wj$ CR (left) and for the $\ttbar$ CR (center). Figure~\ref{fig:wj_FR2_el_nT} (right) shows the $\PW/\PZ$ boson tagging score distributions in the $\Wj$ CR without applying a requirement on the tagger. These distributions have been made before performing the ML fit (pre-fit).  The observed data to MC simulation agreement is within the range of statistical and systematic uncertainties except for the low-$\msd$ regions, as in low tagging score regions in Fig.~\ref{fig:wj_FR2_el_nT} (\cmsRight). A disagreement between data and simulation is observed for smaller values of the tagger score, where the $\Wj$ contribution dominates, while it improves in the regions with higher values of the tagger with increased contributions from processes with genuine $\PV$ bosons from $\ttbar$ and $\WV$ processes. The region with smaller values of $\PW/\PZ$ boson tagging score is dominated by large-$R$ quark or gluon jets mistagged as $\PV$ boson candidate. This mistagging rate is determined during the simultaneous ML fit by freely floating the normalization of the $\Wj$ process after applying the tagger. The normalization and shape of the $\ttbar$ contribution is also determined during the simultaneous ML fit.
The mass peaks from $\PW$ bosons and $\PQt$ quarks are observed at around 90 and 180\GeV, respectively.
The $\msd$ of reconstructed $\PW$ boson jets is shifted by 10\GeV with respect to the corresponding particle mass for the reasons discussed in Section~\ref{eventreco}.

\begin{figure}[!htb]
  \centering
\includegraphics[width=0.5\textwidth]{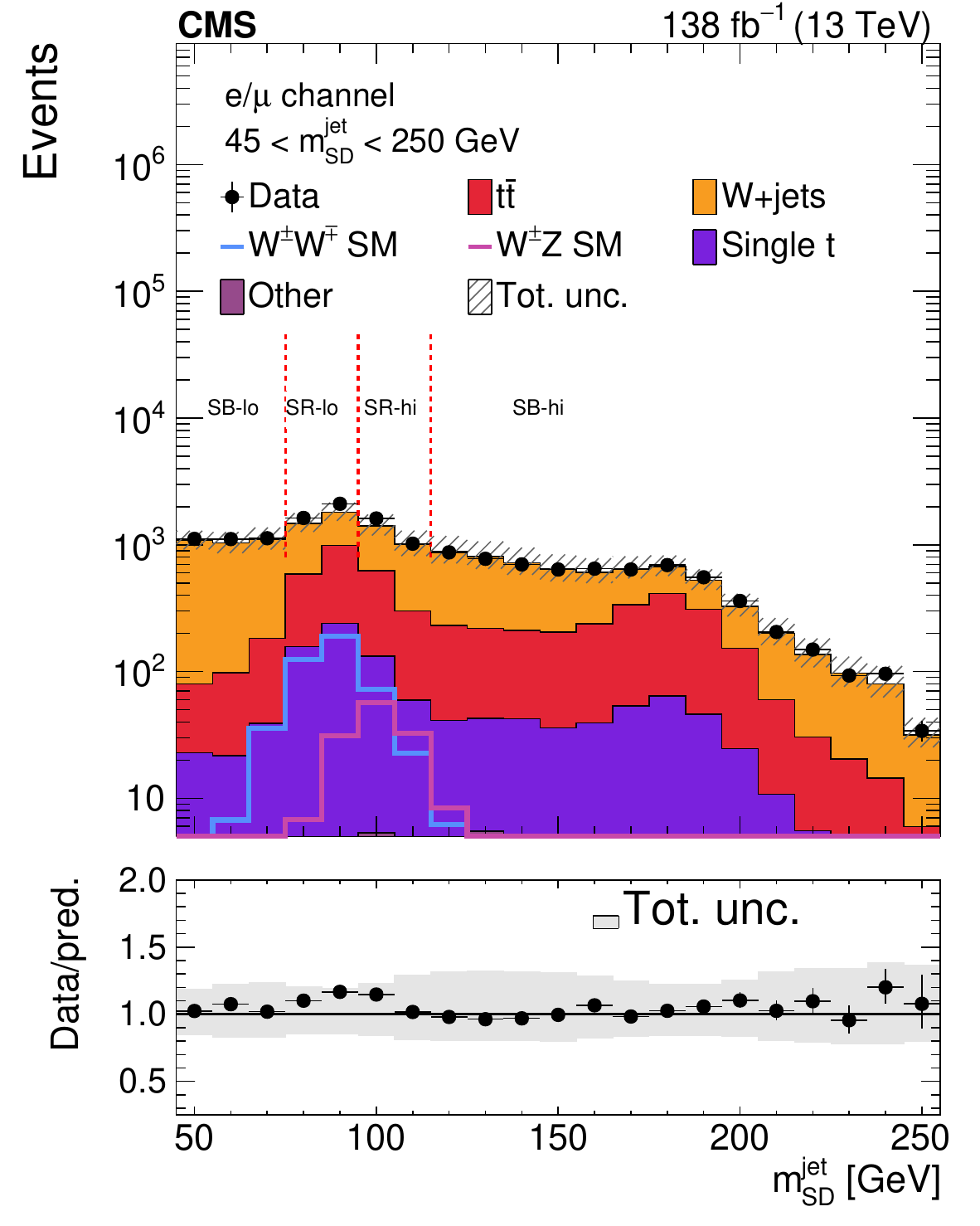}
\caption{Large-R jet $\msd$ in the $\Wj$ CR and the signal region for the combined electron and muon channels and the data-taking years after the ML fit. The SM $\WW$ and $\WZ$ processes are plotted individually as colored lines while rest of the background predictions are shown as filled histograms. The vertical error bars on the data points represent the statistical uncertainty of the data. The uncertainty bands, shown on the shaded histograms and on the ratio plot at the bottom panel, contain both statistical and systematic components on the predictions before the ML fit.}\label{fig:wjpost_FR2_el_nT}
\end{figure}

Figure~\ref{fig:wjpost_FR2_el_nT} shows the $\msd$ distributions for the large-$R$ jet in the $\Wj$ CR and the SR after the ML fit (post-fit) under the background-only hypothesis, \ie, assuming no BSM contributions. The normalizations of different processes have been scaled to those after performing the ML fit to the $\mwv$ distribution as explained earlier. A good agreement between data and simulation is seen, confirming the adequate description of backgrounds in the analysis phase space region. 
The mass peaks from $\PW$ and $\PZ$ bosons in SM $\WV$ production are reconstructed at $\msd$ of around 90 and 110\GeV, respectively, at the center of the $\WW$-sensitive region ($75<\msd<95\GeV$) and the $\WZ$-sensitive region ($95<\msd<115\GeV$). 

Table~\ref{tab:yields} shows the corresponding event yields for SM backgrounds and data in two SRs for electron and muon channels.
The reported post-fit uncertainties include both statistical and systematical components. In this fit the SM $\WV$ cross section is a free parameter and no contribution from EFT signal is included. Overall, we observe consistency with the SM prediction within the range of uncertainties after the fit including a broad range of masses from 950\GeV to $\approx$1.7\TeV where the contribution of EFT signal is expected to be negligible within current constraints. The SM $\WV$ cross section is constrained to less than 30\% uncertainty, which includes both statistical and systematical components, within the phase space of this analysis. This corresponds to a yield that is $0.32\pm0.1$ times the NLO SM prediction. This factor is of similar magnitude to the inverse of the LO-to-NLO QCD $K$-factor included in the SM prediction, thus bringing the post-fit $\WV$ background yield closer to the LO prediction including one extra jet. A dedicated measurement of the differential cross section of SM $\WV$ production over a wide range of vector boson transverse momenta would help clarify this discrepancy. The measurement is limited by statistical uncertainties in data.
\begin{table*}[!htbp]
\centering
\topcaption{\label{tab:yields}
Event yields from SM processes and observed data events in the low and high SRs for the electron and muon channels. The combination of the statistical and systematic uncertainties is indicated. The event yields are shown with their best fit (post-fit) normalizations from the simultaneous fit to the data, assuming no signal, \ie, for the SM case. The contributions from SM $\WW$ and $\WZ$ processes are grouped under the SM $\WV$ category.}
\begin{tabular}{l
r@{\,${}\pm{}$\,}l
r@{\,${}\pm{}$\,}l
r@{\,${}\pm{}$\,}l
r@{\,${}\pm{}$\,}l}
\multicolumn{1}{c}{\multirow{3}{*}{Process}} &
\multicolumn{6}{c}{Signal region} &
\multicolumn{2}{c}{} \\
\multicolumn{1}{c}{} &
\multicolumn{2}{c}{Low} &
\multicolumn{2}{c}{High} &
\multicolumn{2}{c}{Low} &
\multicolumn{2}{c}{High} \\
\multicolumn{1}{c}{} &
\multicolumn{4}{c}{electron} &
\multicolumn{4}{c}{muon} \\
\hline\\
SM \WV   &  172 &    9 &   81 &    4 &  183 &	10 &   81 &   5 \\
$\Wj$    &  812 &   13 &  708 &   12 &  952 &	15 &  822 &  13 \\
$\ttbar$ &  571 &    8 &  344 &    6 &  612 &	 8 &  387 &   6 \\
Single t &  187 &    6 &   82 &    4 &  209 &	 7 &   97 &   4 \\
Other    &  2.8 & 0.04 &  4.1 & 0.02 &  3.2 & 0.02 &  6.2 & 0.03 \\
Total SM & 1746 &   18 & 1221 &   14 & 1961 &	19 & 1394 &   14 \\
[\cmsTabSkip]
\multicolumn{1}{l}{Data}
& \multicolumn{2}{c}{1748}
& \multicolumn{2}{c}{1220}
& \multicolumn{2}{c}{1984}
& \multicolumn{2}{c}{1412} \\
\end{tabular}
\end{table*}

\section{Signal extraction and results}\label{sec:results}
We set constraints on the WCs in an ML fit
to data using the \textsc{Combine} statistical analysis tool~\cite{CMS:2024onh}, based on the \textsc{RooFit}~\cite{Verkerke:2003ir} and \textsc{RooStats}~\cite{Moneta:2010pm} frameworks, using the asymptotic approximation~\cite{Cowan:2010js,CMS-NOTE-2011-005} for the distribution of the profile likelihood ratio test statistic.
A simultaneous binned ML fit is carried out to the $\mWV$ distributions in the SR and the background CRs.
In the fit, one or two WCs are treated as parameter of interest, while the remaining WCs are set to zero.
The likelihood is built from Poisson probabilities to observe the per-bin event yields in data,
given the signal and background predictions.
We combine the likelihoods of all years of data taking during the fit.
Sources of systematic uncertainties are modeled with nuisance parameters.
Shape uncertainties are implemented using a Gaussian distribution and
rate uncertainties are modeled with a log-normal distribution.
Confidence intervals at 68\% and 95\% on the WCs are set by requiring the negative log-likelihood ratio, $-2\Delta \ln L$, to be less than 0.99 and 3.84, respectively, for limits on single parameters, and less than 2.28 and 5.99, respectively, for two-parameter scans.
Corresponding expected confidence intervals are computed using an Asimov data set~\cite{Cowan:2010js} (background-only prediction) where the $\WV$ cross section is set to its post-fit value from the background-only fit.

\begin{figure*}[htbp!]
\centering
\includegraphics[width=0.45\textwidth]{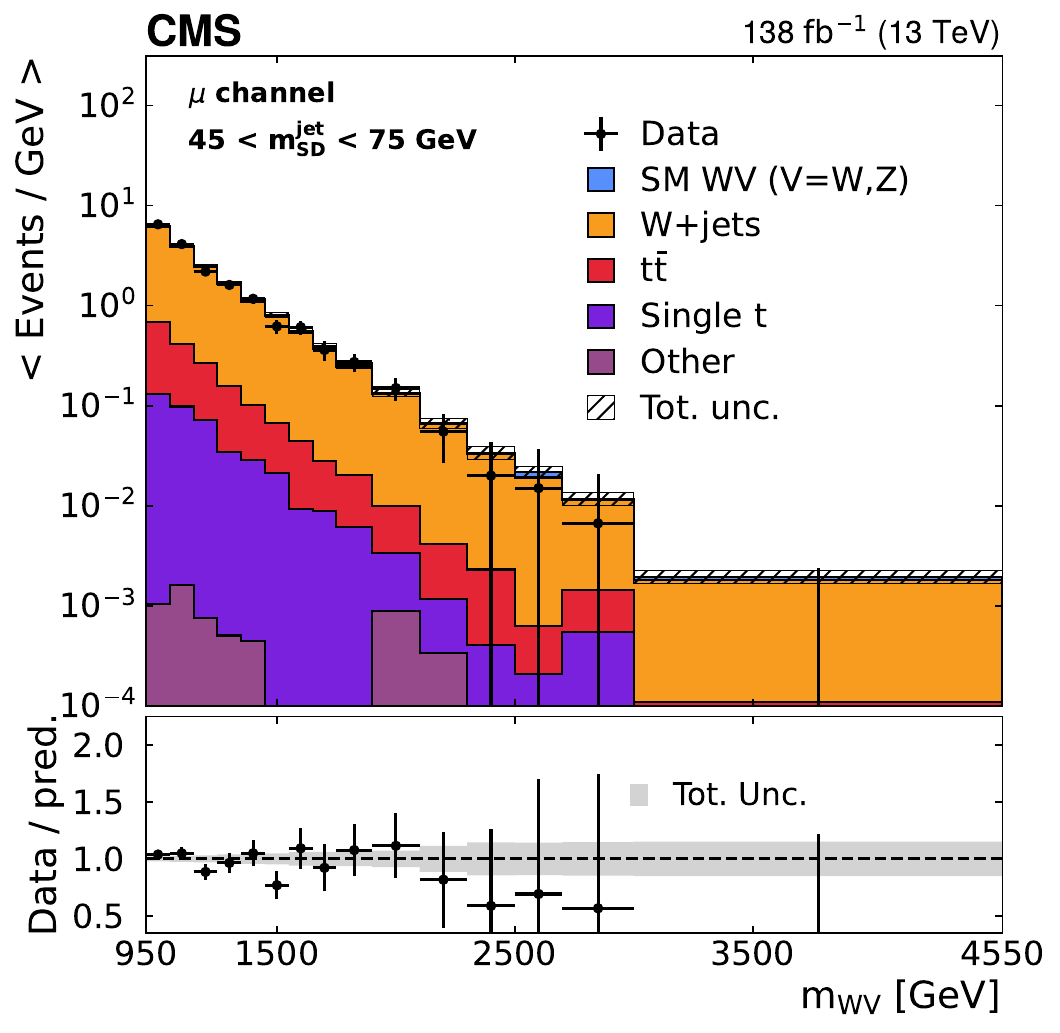}
\includegraphics[width=0.45\textwidth]{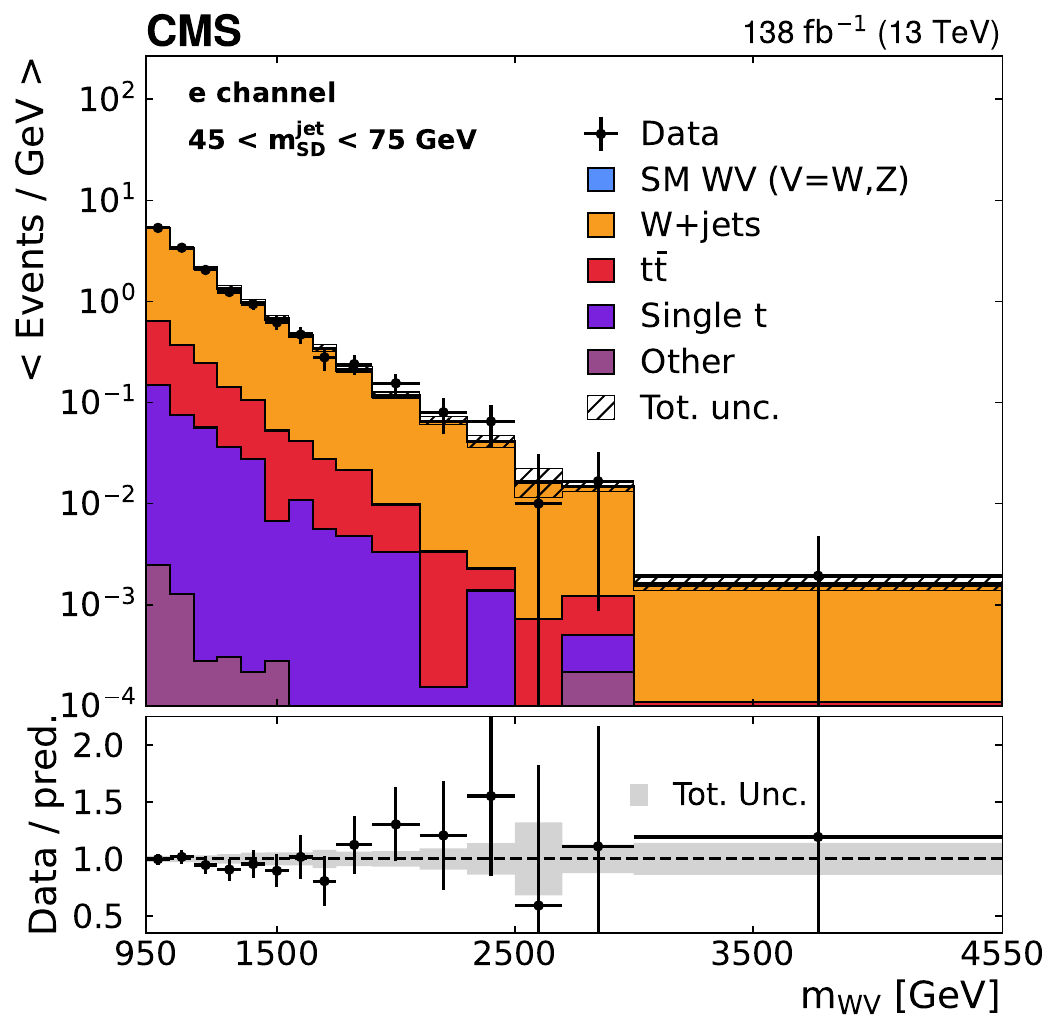}\\
\includegraphics[width=0.45\textwidth]{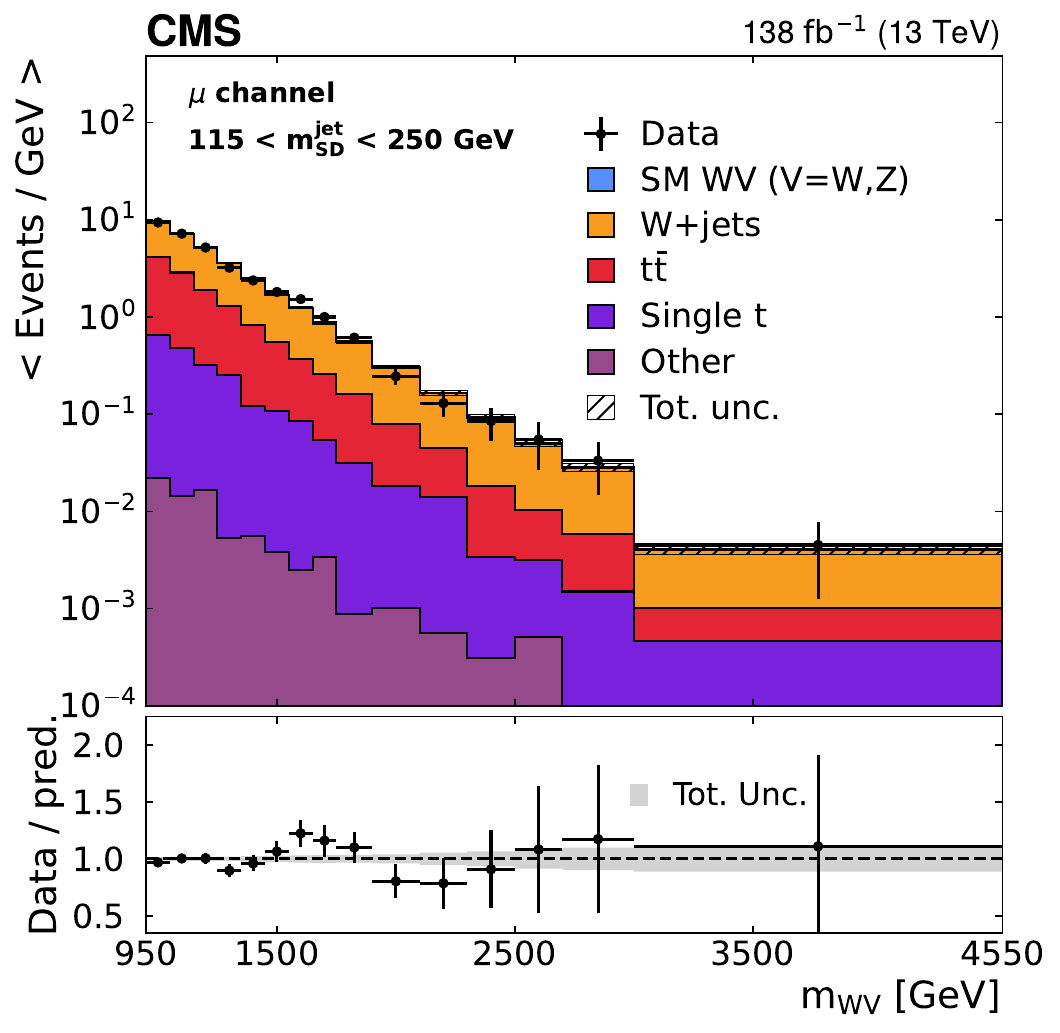}
\includegraphics[width=0.45\textwidth]{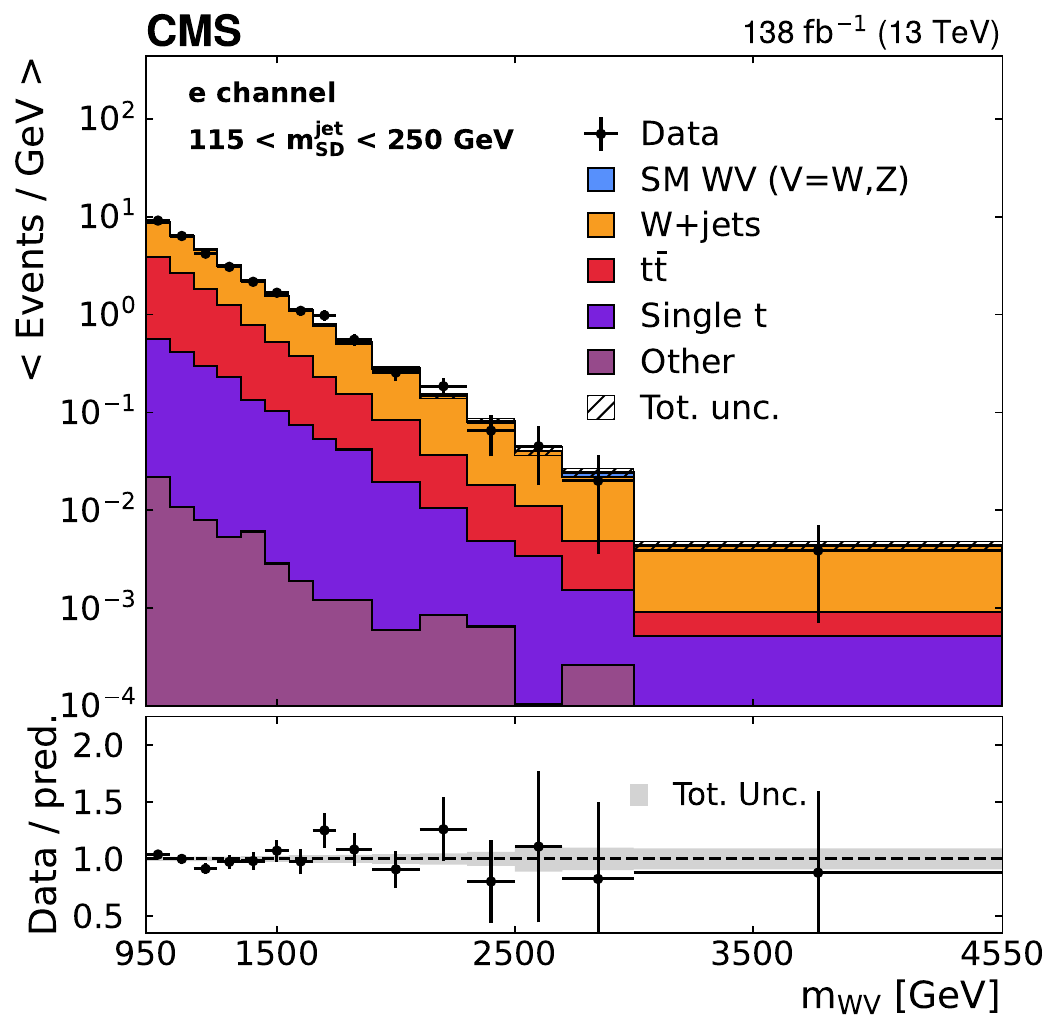}\\
\includegraphics[width=0.45\textwidth]{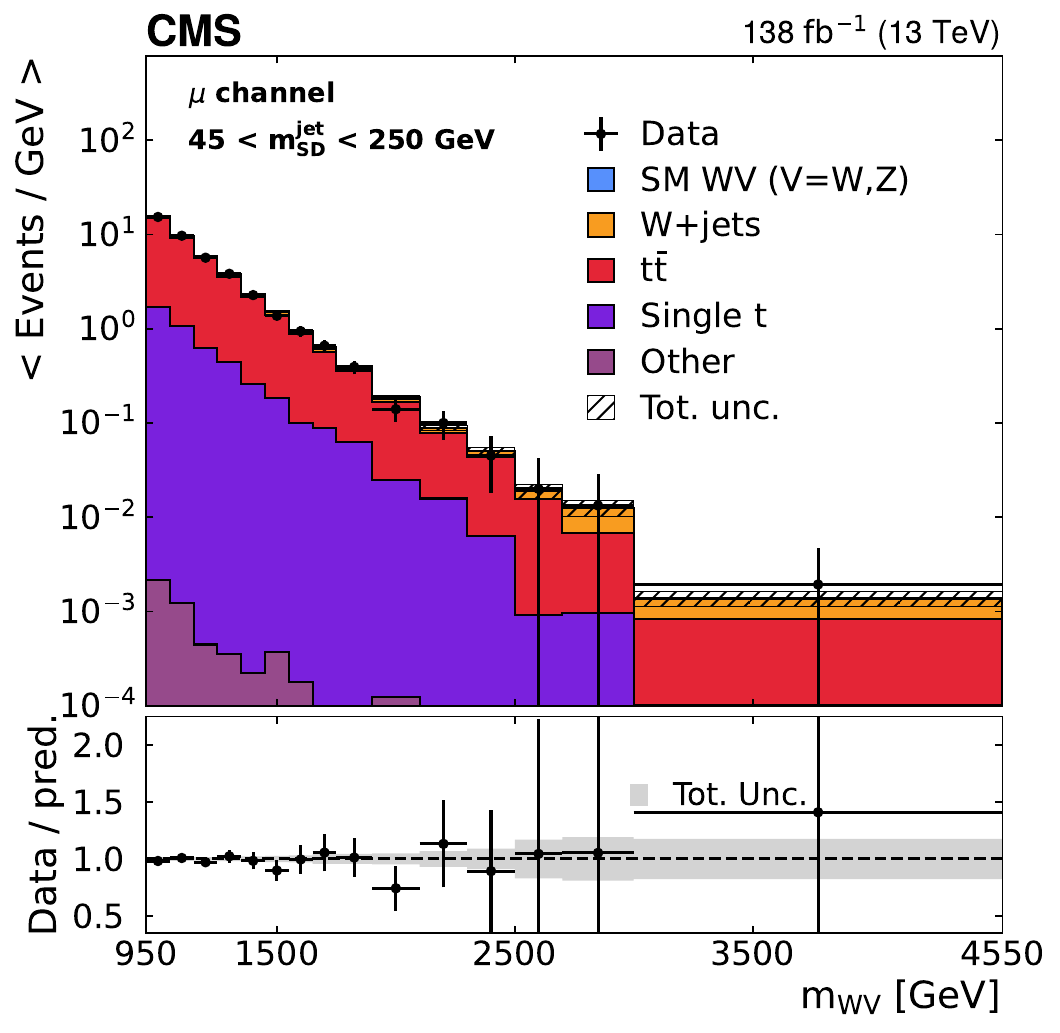}
\includegraphics[width=0.45\textwidth]{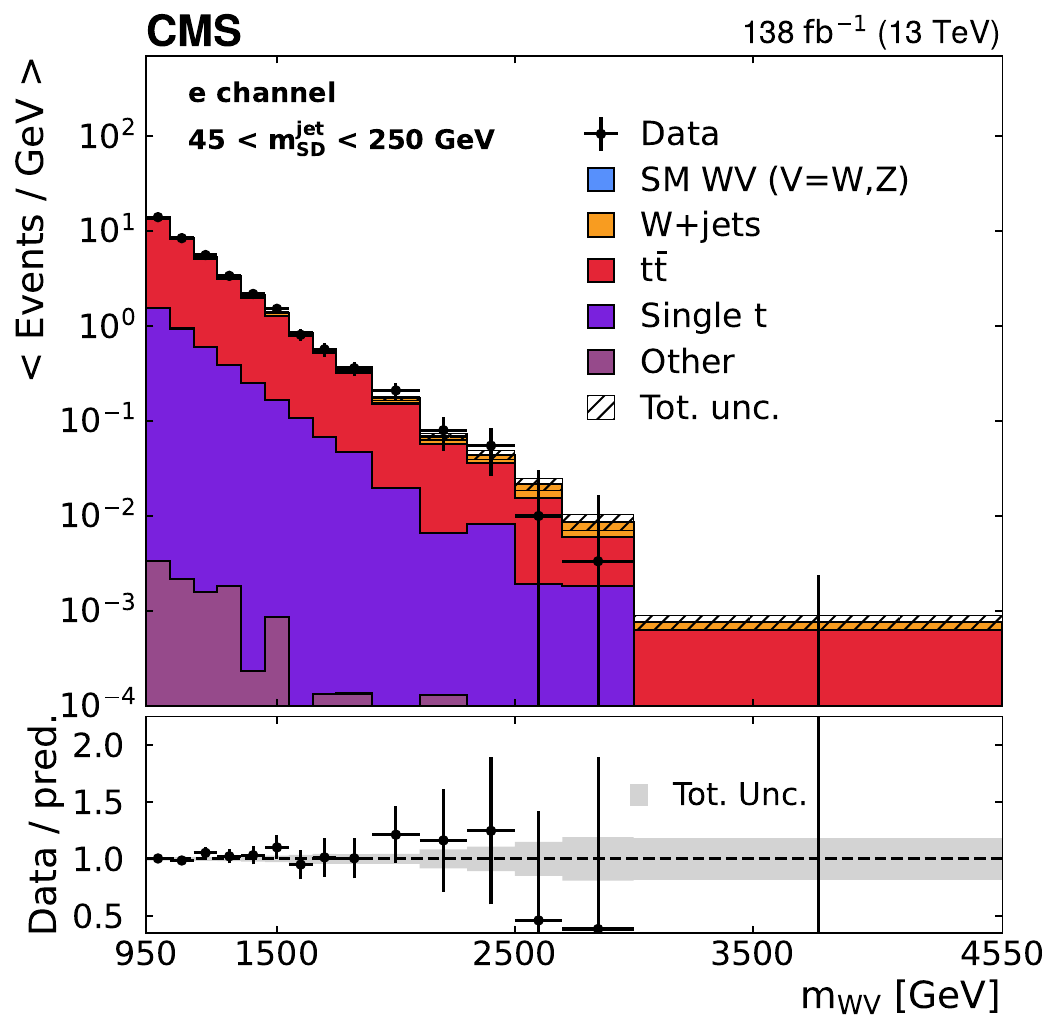}
\caption{The $\mWV$ distribution in the $\Wj$ low (\cmsLeft) and high (center) sidebands and $\ttbar$ (\cmsRight) control regions in the muon (left) and electron (right) channels after the ML fit considering the SM-only scenario (background-only fit). The vertical error bars on the data points represent the statistical uncertainty of the data. The lower panels show the ratio of the data to the postfit signal plus background prediction. The uncertainty bands, shown on the shaded histograms and on the ratio plot at the bottom panel, contain both statistical and systematic components on the predictions after the ML fit.}\label{fig:bg_FR2_el}
\end{figure*}

\begin{figure*}[htb!]
\centering
\includegraphics[width=0.45\textwidth]{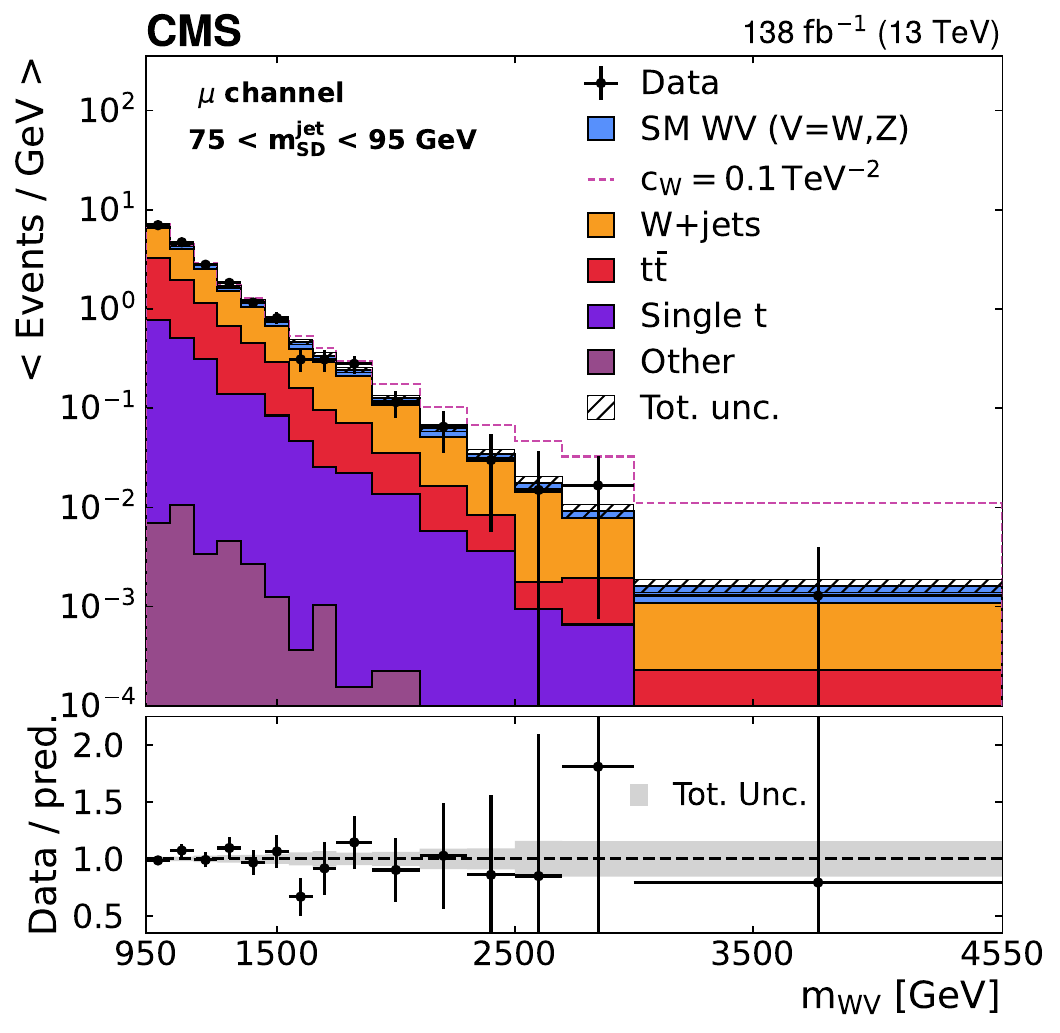}
\includegraphics[width=0.45\textwidth]{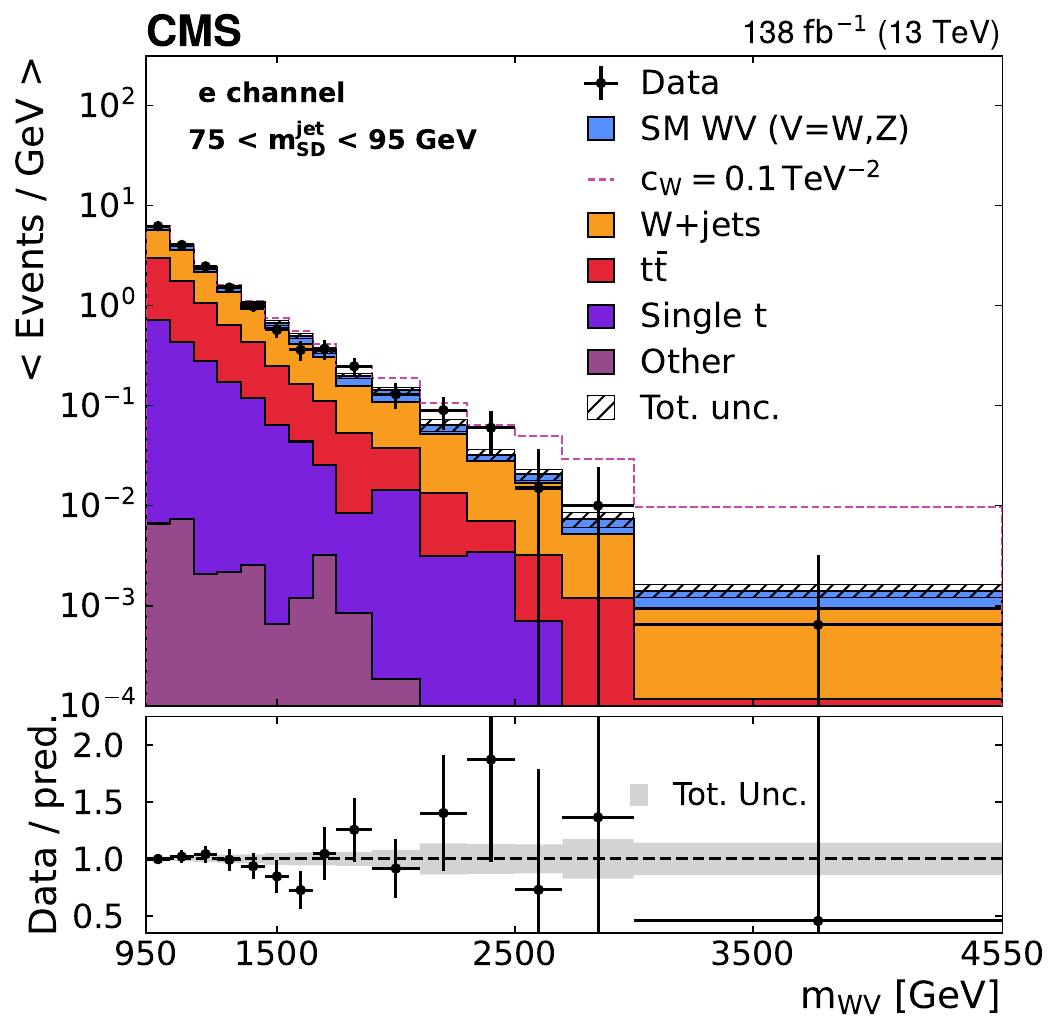}\\
\includegraphics[width=0.45\textwidth]{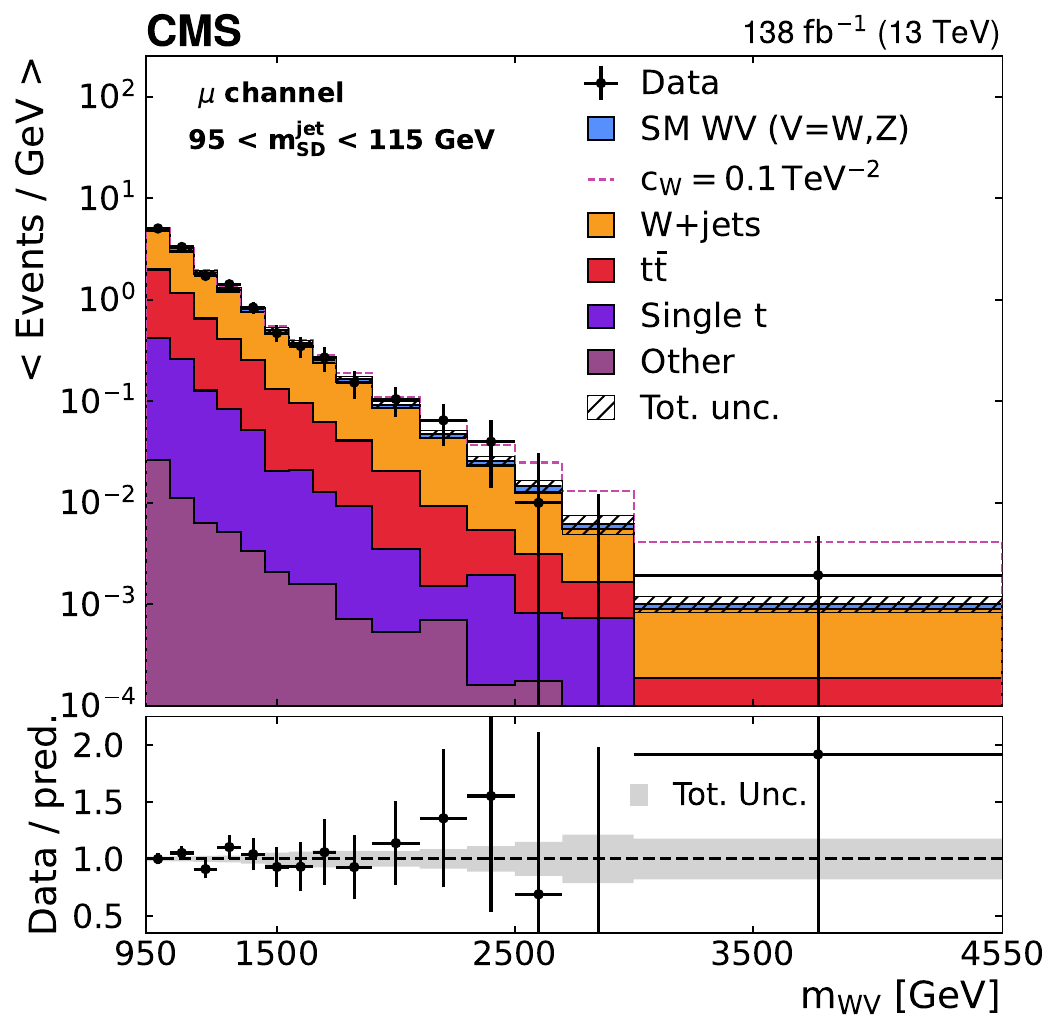}
\includegraphics[width=0.45\textwidth]{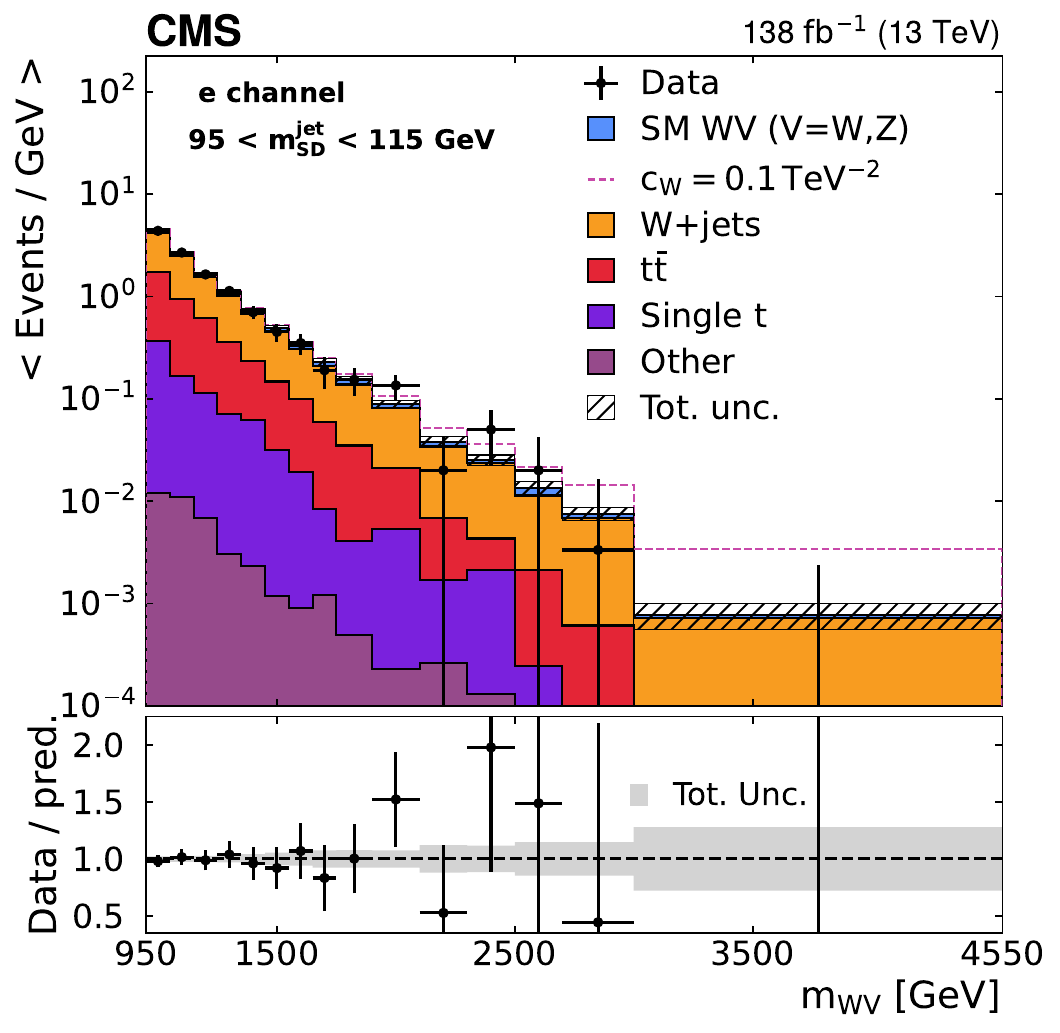}
\caption{The $\mWV$ distribution in the signal region low (\cmsLeft) and high (\cmsRight) in the muon (left) and electron (right) channels after the ML fit assuming no signal. The prefit signal contribution from the summed linear and quadratic terms corresponding to $\cw = 0.1 \TeV^{-2}$ are also shown. The vertical error bars on the data points represent the statistical uncertainty of the data. The lower panels show the ratio of the data to the postfit signal plus background prediction. The uncertainty bands, shown on the shaded histograms and on the ratio plot at the bottom panel, contain both statistical and systematic components on the predictions after the ML fit.}\label{fig:sig_FR2_el}
\end{figure*}

As an example, we present one case that illustrates the typical $\mWV$ distribution after the ML fit, observed likelihood profiles, and the impact of uncertainties.
Figure~\ref{fig:bg_FR2_el} shows the $\mWV$ distributions in the $\Wj$ and $\ttbar$ CRs after the ML fit under the SM-only scenario \ie no signal from BSM (background-only fit), and Fig.~\ref{fig:sig_FR2_el} the corresponding distributions in the SR.
Expected contributions from signal are shown as dotted lines corresponding to the value of the WC $\cw=0.1\TeV^{-2}$ in the SR.
The data in both CRs are well described by the background prediction with negligible contribution from SM \WV processes.
In the SR, the signal is expected to lead to an enhancement of events in the tail of the $\mwv$ distribution as indicated by the expected signal contribution.

\begin{figure}[ht!]
\centering
\includegraphics[width=0.45\textwidth]{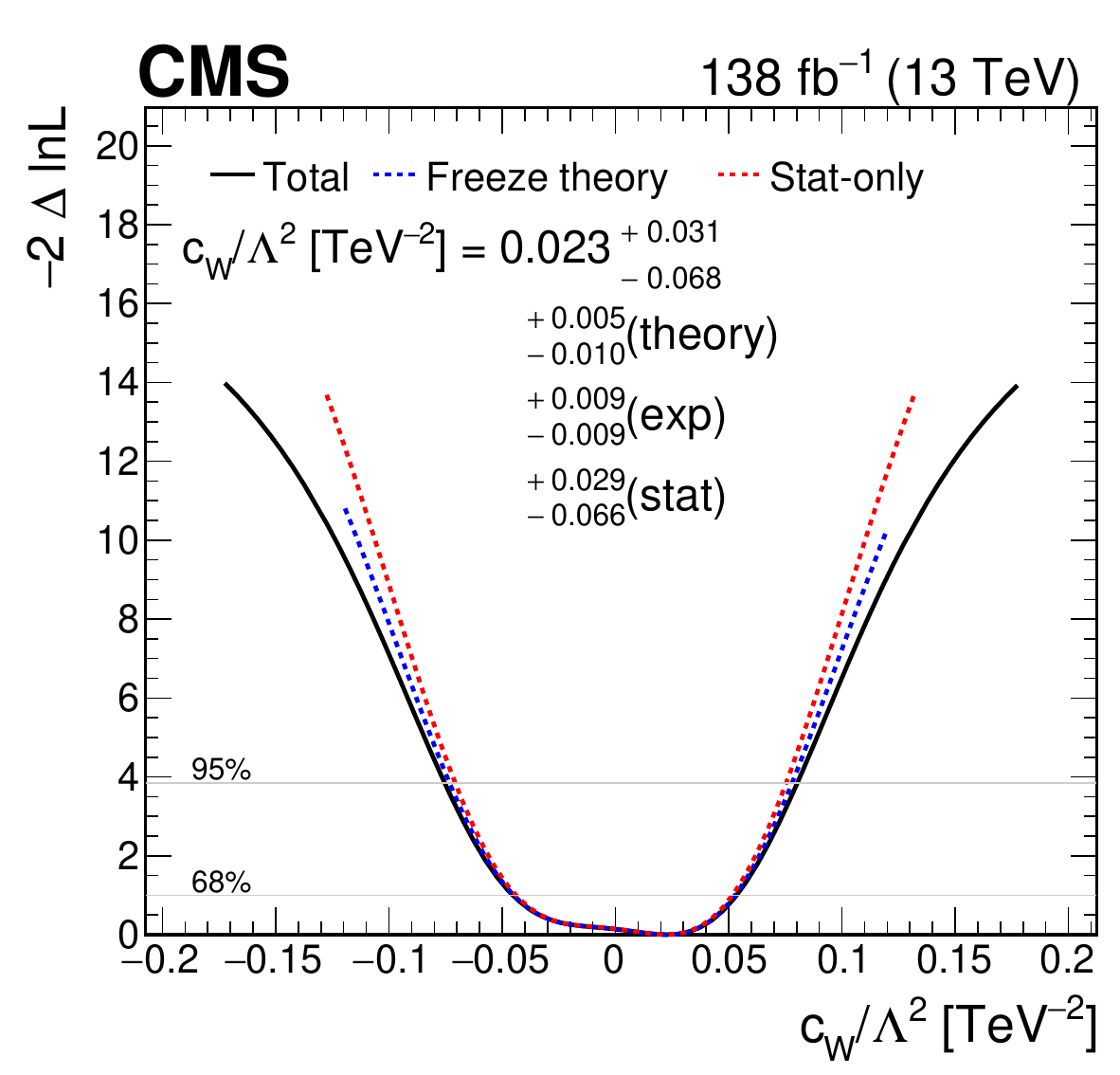}
\caption{Profile likelihood scan of the WC $\cw$ obtained from a simultaneous fit to data in the SRs and background CRs for the case when all other WCs are fixed to zero. Both linear and quadratic contributions in the EFT expansion are included in the fit. The horizontal dashed lines correspond to the 68\% and 95\% confidence levels. Different profile likelihood curves correspond to cases where: all systematic uncertainty nuisance parameters are included and varied in the fit, theory uncertainty parameters are fixed to their best fit values, and where all nuisance parameters are fixed (statistical only).}\label{fig:nll_scan_cW}
\end{figure}

Figure~\ref{fig:nll_scan_cW} shows the likelihood profile for the WC $\cw$ obtained from the simultaneous fit to data in the SRs and background CRs for the case when all other WCs are fixed to zero. The total uncertainty in $\cw$ is decomposed into statistical, experimental, and theoretical components, and is dominated by the statistical component. Asymmetries in the likelihood may arise when there is both a linear and quadratic dependence on the WC in $\abs{\mathcal{M}}^{2}$ that exhibit different dependence on the sign of the WC.
As a second-order effect, asymmetries may arise from systematic uncertainties affecting the SM $\WV$ yield.

The resulting limits on the WCs in the HISZ model are summarized in Table~\ref{tab:limitsEFT}.
\begin{table*}[!htbp]
 \centering
\topcaption{\label{tab:limitsEFT}
Expected and observed individual limits on the WCs, from HISZ basis as 68\% and 95\% confidence intervals.}
\begin{tabular}{ccccc}
 \multicolumn{1}{c}{\multirow{3}{*}{WC}} & \multicolumn{4}{c}{Limits ($\TeV^{-2}$)}                                  \\
\multicolumn{1}{c}{}                          & \multicolumn{2}{c}{Expected} & \multicolumn{2}{c}{Observed} \\
\multicolumn{1}{c}{}                         & 68\%            & 95\%            & 68\%           & 95\%           \\  [\cmsTabSkip] \hline 
\noalign{\vskip 0.5mm} 
  $c_{\PB}^\mathrm{HISZ}/\Lambda^2$             &   [-3.09,\,2.92]  &  [-4.69,\,4.49]   &  [-2.81,\,3.32]  &  [-5.18,\,5.39]\\  [\cmsTabSkip]
  $c_{\PW\PW\PW}^\mathrm{HISZ}/\Lambda^2$       &   [-0.55,\,0.55]  &  [-0.81,\,0.82]   &  [-0.58,\,0.55]  &  [-0.93,\,0.93]\\  [\cmsTabSkip]
  $c_{\PW}^\mathrm{HISZ}/\Lambda^2$             &   [-1.15,\,0.70]  &  [-1.66,\,1.14]   &  [-0.56,\,1.02]  &  [-1.44,\,1.78]\\  [\cmsTabSkip]
  $c_{\widetilde{W}WW}^\mathrm{HISZ}/\Lambda^2$ &   [-0.26,\,0.27]  &  [-0.40,\,0.40]   &  [-0.13,\,0.33]  &  [-0.37,\,0.49]\\  [\cmsTabSkip]
  $c_{\widetilde{W}}^\mathrm{HISZ}/\Lambda^2$   &   [-1.66,\,1.12]  &  [-2.37,\,1.79]   &  [-1.31,\,1.24]  &  [-2.24,\,2.13] \\  [\cmsTabSkip]
\end{tabular}
 \end{table*}

The limits in the HISZ basis significantly improve by 30--70\% over previous results with $-8.78<c_{\PB}^\mathrm{HISZ}/\Lambda^{2}<8.54\TeV^{-2}$, $-1.58<c_{\PW\PW\PW}^\mathrm{HISZ}/\Lambda^{2}<1.59\TeV^{-2}$, and $-2.00<c_{\PW}^\mathrm{HISZ}/\Lambda^{2}<2.65\TeV^{-2}$ at 95\% confidence level~\cite{CMS:2019ppl}.

\begin{figure}
\centering
\includegraphics[width=\cmsFigWidth]{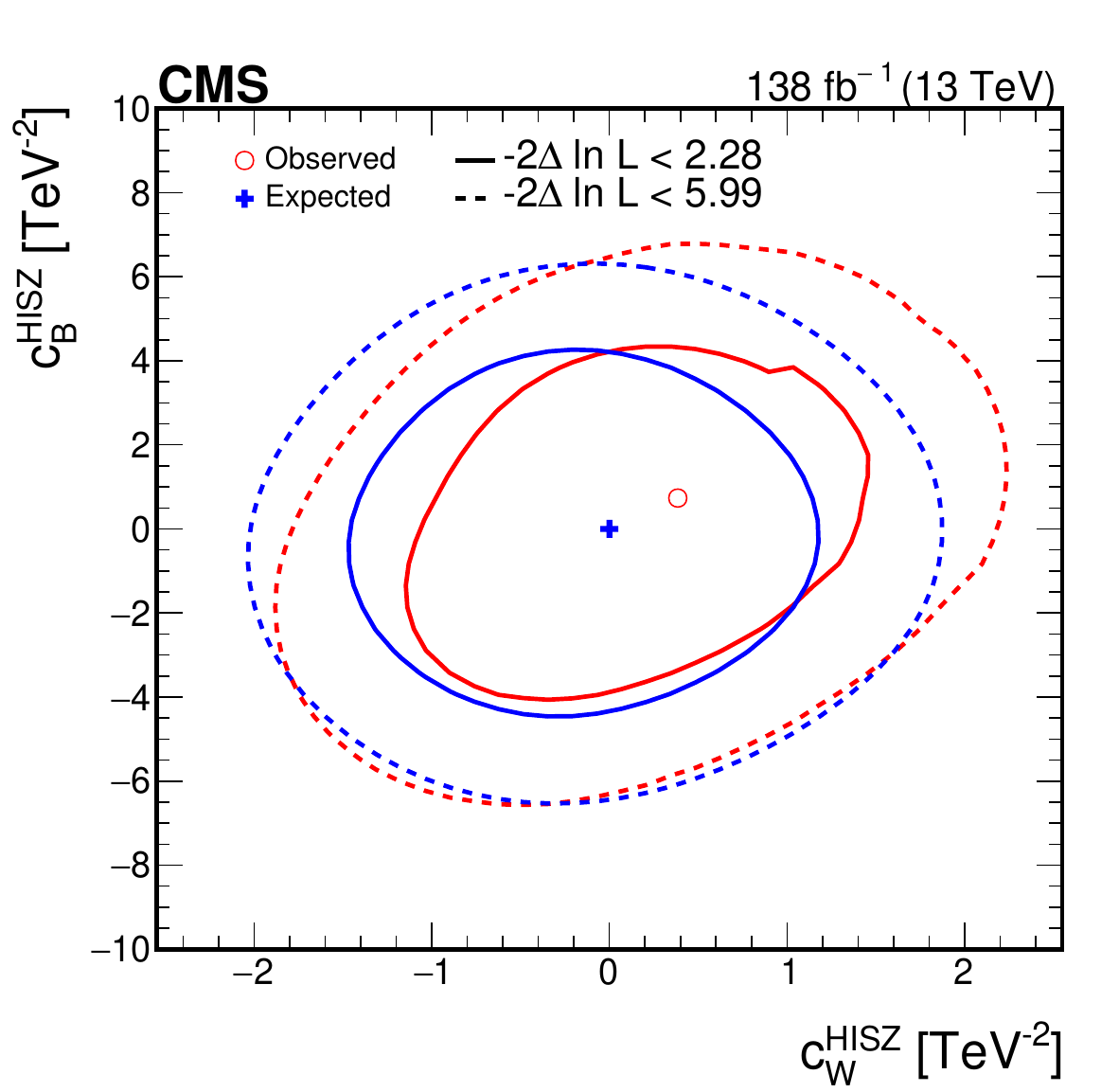}
\includegraphics[width=\cmsFigWidth]{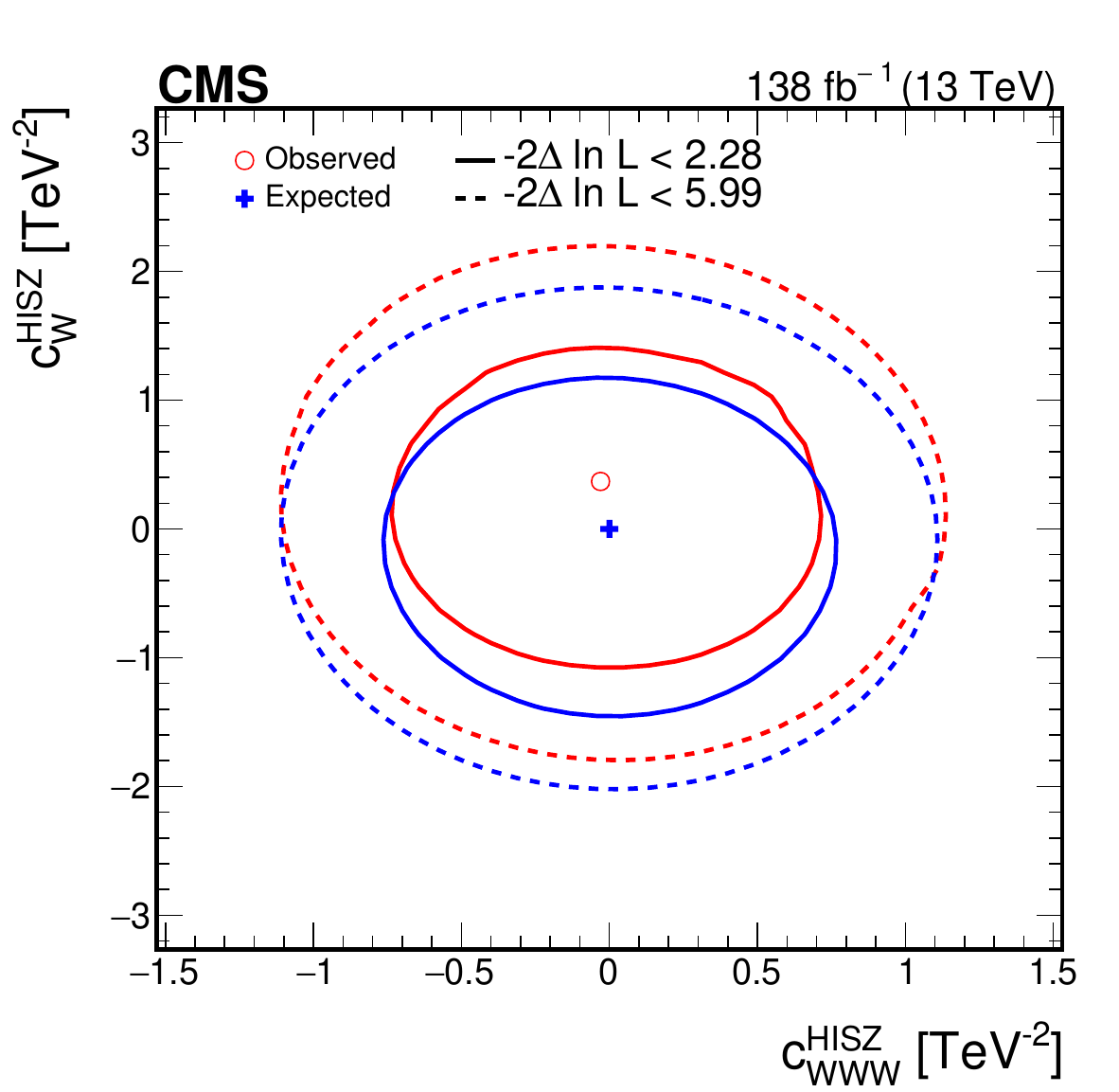}
\includegraphics[width=\cmsFigWidth]{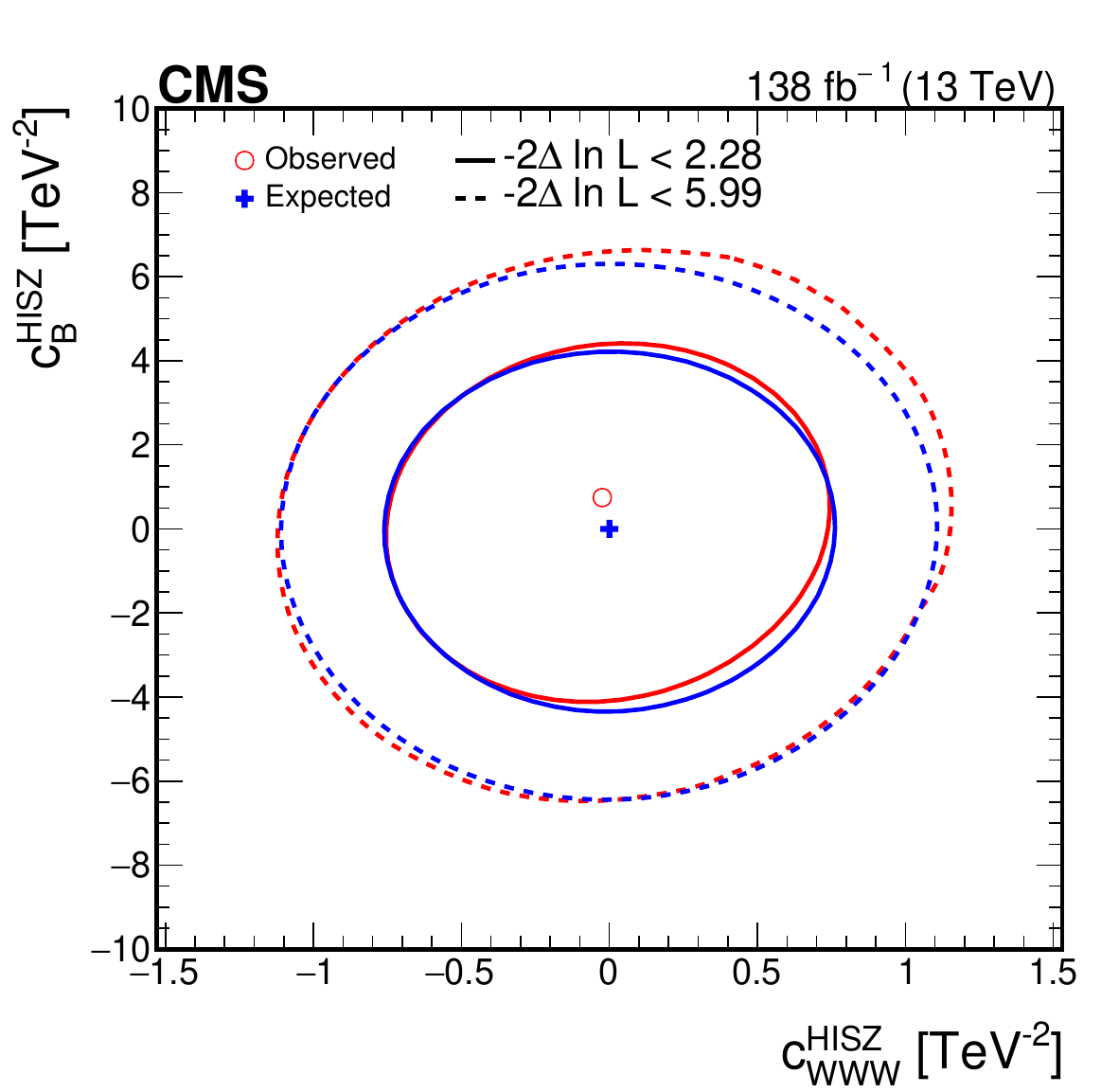}
\caption{Likelihood scans as functions of the pairs of WCs from HISZ model: $c_{\PW}$ and $c_{\PB}$ (\cmsLeft), $c_{\PW\PW\PW}$ and $c_{\PW}$ (center), and $c_{\PW\PW\PW}$ and $c_{\PB}$ (\cmsRight). All WCs that are not scanned are fixed to zero. The best fit value is shown with a marker and the dotted lines correspond to the crossing points of $-2\Delta\ln L$ at 2.28 and 5.99, which correspond to the 68\% and 95\% confidence levels in the asymptotic approximation.}\label{fig:2Dlimits_eft}
\end{figure}

Figure~\ref{fig:2Dlimits_eft} shows two-dimensional likelihood scans for pairs of WCs in the HISZ basis. These do not show any significant linear correlation, as evident from the circular shapes of the contours.

The limits on WCs from the SMEFT model are summarized in Table~\ref{tab:limitsSMEFT}. 

\begin{table*}[!htbp]
  \centering
 \topcaption{\label{tab:limitsSMEFT}
 Expected and observed individual limits on WCs in SMEFT scenarios as 68\% and 95\% confidence intervals.}
 \begin{tabular}{ccccc}
 \multicolumn{1}{c}{\multirow{3}{*}{WC}} & \multicolumn{4}{c}{Limits ($\TeV^{-2}$)}                                  \\
\multicolumn{1}{c}{}                          & \multicolumn{2}{c}{Expected} & \multicolumn{2}{c}{Observed} \\
\multicolumn{1}{c}{}                          & 68\%                 & 95\%                    & 68\%          & 95\%           \\  [\cmsTabSkip] \hline
\noalign{\vskip 0.5mm} 
$\cw/\Lambda^2$                               &        [-0.04,\,0.04]  &         [-0.06,\,0.06]    & [-0.05,\,0.05]  &         [-0.08,\,0.08]   \\
$\cwtil/\Lambda^2$         		      &        [-0.04,\,0.04]  &         [-0.06,\,0.06]    & [-0.05,\,0.05]  &         [-0.08,\,0.08]   \\
$\chwb/\Lambda^2$          		      &        [-1.95,\,1.72]  &         [-2.80,\,2.56]    & [-3.27,\,3.33]  &         [-6.25,\,6.70]   \\
$\chwbtil/\Lambda^2$       		      &        [-1.86,\,1.87]  &         [-2.71,\,2.70]    & [-2.58,\,3.22]  &         [-5.27,\,6.07]   \\
$\chd/\Lambda^2$           		      &        [-0.10,\,0.11]  &         [-0.15,\,0.16]    & [-0.10,\,0.10]  &         [-0.17,\,0.17]   \\
$\chu/\Lambda^2$           		      &        [-0.08,\,0.07]  &         [-0.12,\,0.11]    & [-0.08,\,0.07]  &         [-0.13,\,0.12]   \\ [\cmsTabSkip]
$\chj{1}/\Lambda^2$        		      &        [-0.07,\,0.06]  &         [-0.10,\,0.09]    & [-0.06,\,0.06]  &         [-0.10,\,0.10]   \\               [\cmsTabSkip]        
$\chj{3}/\Lambda^2$        		      &        [-0.06,\,0.03]  &         [-0.08,\,0.05]    & [-0.05,\,0.05]  &         [-0.08,\,0.08]   \\ [\cmsTabSkip]
$\cjd{1}/\Lambda^2$        		      &        [-1.03,\,1.02]  &         [-1.56,\,1.52]    & [-0.91,\,0.86]  &         [-1.58,\,1.53]    \\ [\cmsTabSkip]
$\cju{1}/\Lambda^2$        		      &        [-0.94,\,0.94]  &         [-1.40,\,1.38]    & [-0.53,\,0.52]  &         [-1.01,\,1.00]   \\ [\cmsTabSkip]
$\cjj{1}{1}/\Lambda^2$     		      &        [-0.20,\,0.18]  &         [-0.30,\,0.28]    & [-0.15,\,0.15]  &         [-0.26,\,0.26]    \\ [\cmsTabSkip]
$\cjj{1}{8}/\Lambda^2$     		      &        [-0.43,\,0.45]  &         [-0.67,\,0.68]    & [-0.32,\,0.36]  &         [-0.59,\,0.62]    \\ [\cmsTabSkip]
$\cjj{3}{1}/\Lambda^2$     		      &        [-0.04,\,0.04]  &         [-0.06,\,0.06]    & [-0.03,\,0.03]  &         [-0.05,\,0.05]    \\ [\cmsTabSkip]
$\cjj{3}{8}/\Lambda^2$     		      &        [-0.04,\,0.04]  &         [-0.06,\,0.06]    & [-0.03,\,0.03]  &         [-0.05,\,0.05]   \\ [\cmsTabSkip]
$\clj{1}/\Lambda^2$        		      &        [-0.27,\,0.31]  &         [-0.45,\,0.51]    & [-0.57,\,-0.21] &         [-0.76,\,0.65]    \\ [\cmsTabSkip]
$\clj{3}/\Lambda^2$        		      &        [-0.32,\,0.27]  &         [-0.49,\,0.43]    & [0.26,\,0.61]   &         [-0.87,\,0.83]   \\ [\cmsTabSkip]
$\clu/\Lambda^2$           		      &        [-0.42,\,0.43]  &         [-0.64,\,0.64]    & [-0.45,\,0.37]  &         [-0.70,\,0.65]   \\ [\cmsTabSkip]
$\cld/\Lambda^2$           		      &        [-0.54,\,0.53]  &         [-0.83,\,0.80]    & [-0.89,\,-0.21] $\cup$ [0.22,\,0.94] &         [-1.24,\,1.31]    \\
\end{tabular}
 \end{table*}

\begin{figure}[!ht]
  \centering
  \includegraphics[width=.5\textwidth]{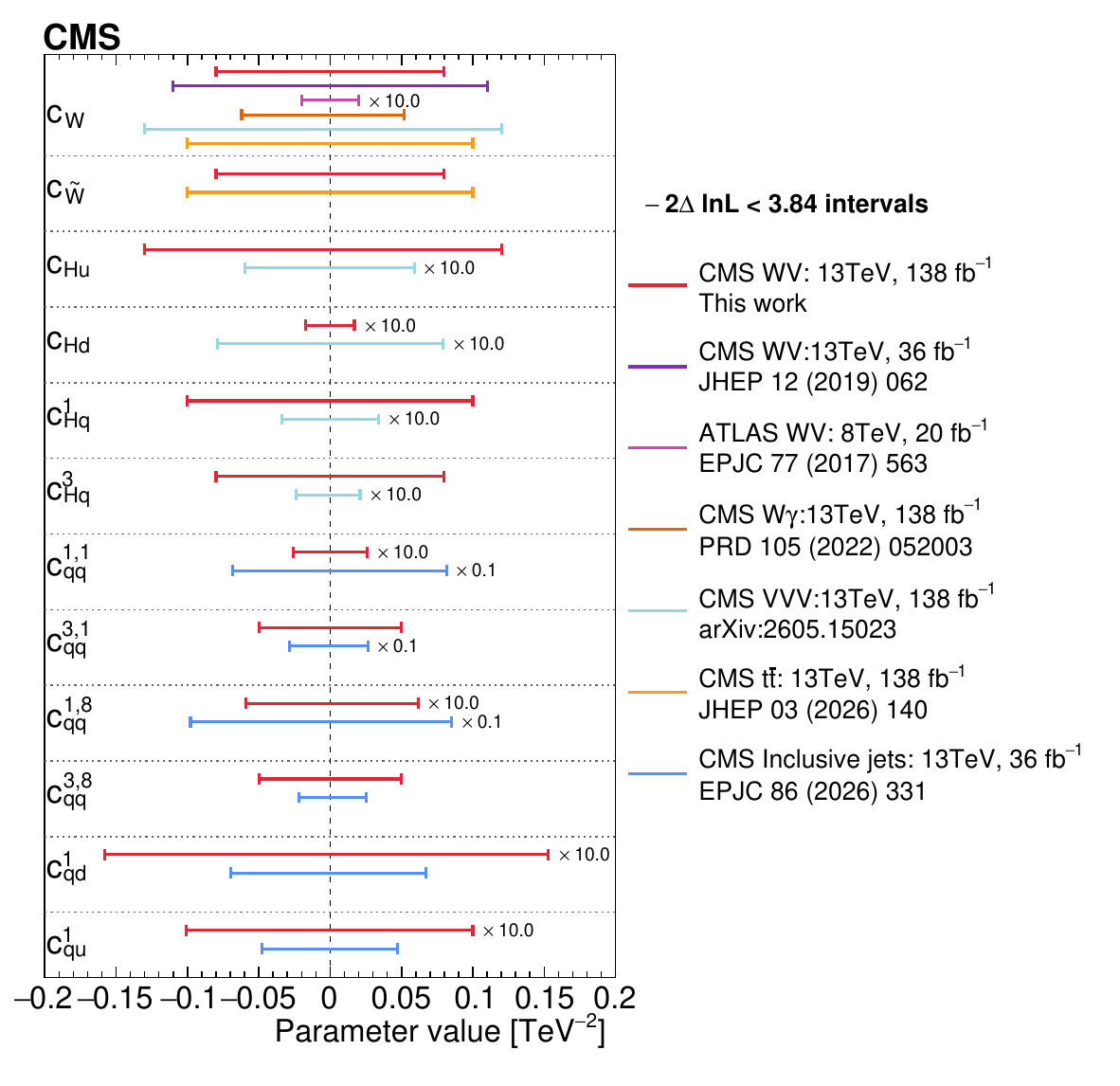}
  \caption{Comparison of observed limits on different WCs at 95\% confidence level with existing analyses at different center-of-mass energies and different final states~\cite{CMS-PAS-TOP-23-009,SMP-24-003,CMS-PAS-SMP-24-017,Aad:2016wpd,Sirunyan:2017bey}. The limits from some of the analyses are scaled down or up so as to fit on the same scale as for the best constraints. The corresponding scale factor is mentioned, next to the limits, which should be multiplied with the plotted limits to extract the actual constraints.}\label{fig:eft_comp}
\end{figure}

Figure~\ref{fig:eft_comp} shows a comparison of obtained limits on different WCs compared with those obtained from other analyses at different center-of-mass energies and different final states.
This analysis provides the strongest constraints to date on the SMEFT operators affecting triple gauge boson couplings.
The inclusive jet measurements provide the most stringent bounds to date on the four light-quark operators.

\begin{figure*}[!htb]
  \centering
  \includegraphics[width=.8\textwidth]{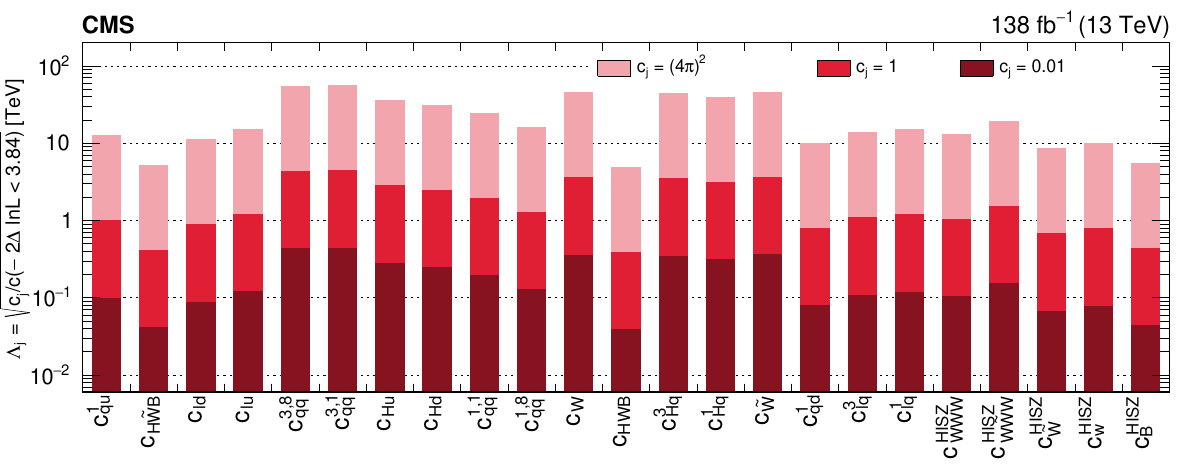}
  \caption{Summary of the lower limits on the energy scales $\Lambda_{j}$, obtained from the observed limits on the WCs at 95\% confidence interval, for the indicated values of the $c_{j}$.}\label{fig:sum}
\end{figure*}

The observed limits on each WC are used to derive lower limits of the BSM physics energy scale ($\Lambda$).
These limits are obtained via $\Lambda = \sqrt{\smash[b]{c_{j}/c(q=3.84)}}$, where $c_{j}$ is a fixed value and $c(q=3.84)$ corresponds to the value of the WC at $q = -2\Delta\ln L = 3.84$.
The WC can take a broad range of possible values, depending on whether it arises from weakly or strongly coupled BSM physics, at tree level or at loop level. For the loop expansion to converge,
the condition $c_j < (4\pi)^2$ must be satisfied~\cite{Degrande_2013}.
In line with the conventions used in other EFT interpretations, such as in Ref.~\cite{Ellis_2021}, we display the lower limits on the scale $\Lambda_j$ for $c_j = 0.01$, 1, and $(4\pi)^2$.
The observed constraints on the WCs at 95\% confidence levels are translated into lower limits on the energy scale of BSM physics as shown in Fig~\ref{fig:sum}.

\begin{figure*}[!htb]
\centering
\includegraphics[width=0.45\textwidth]{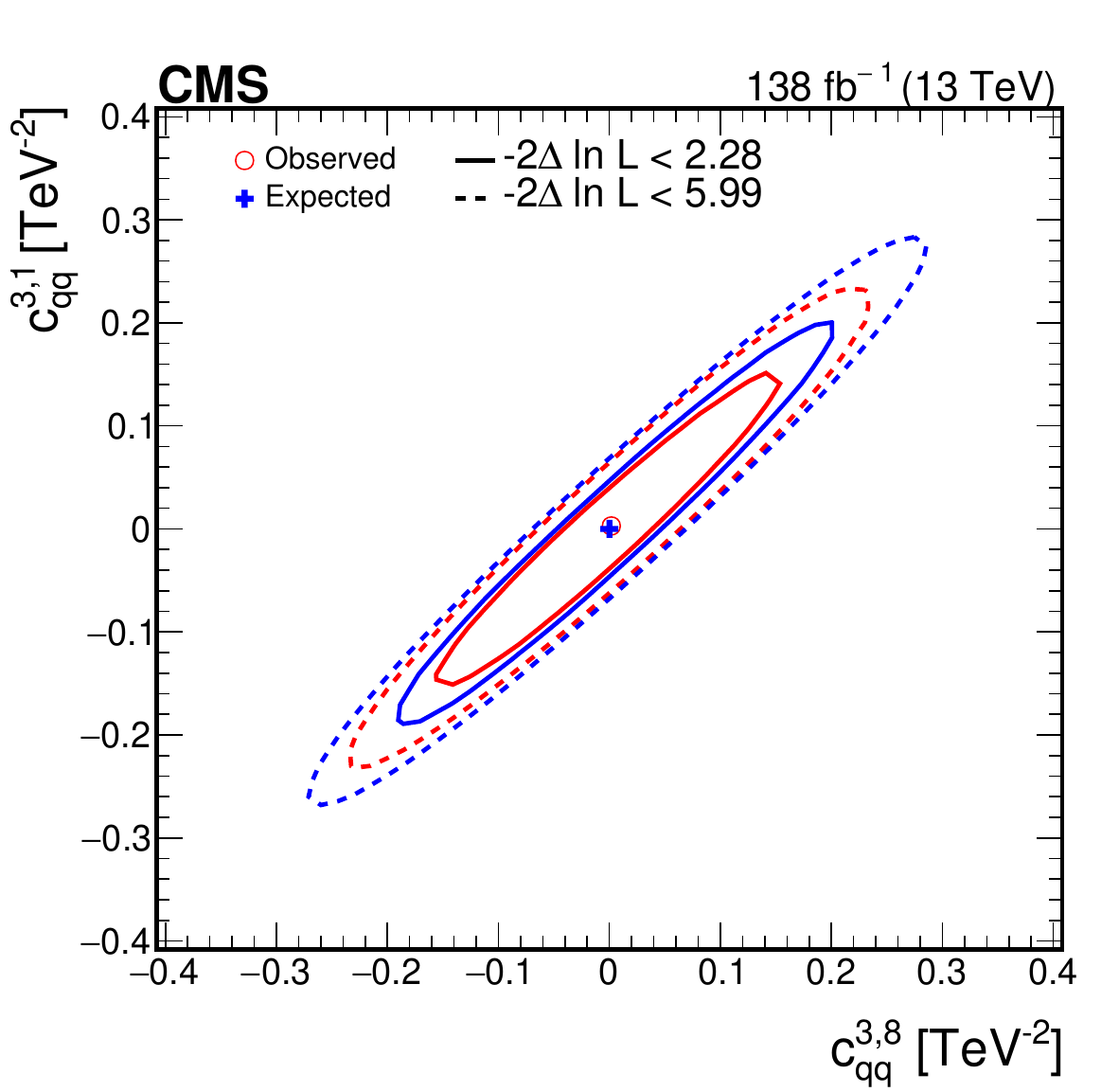}
\includegraphics[width=0.45\textwidth]{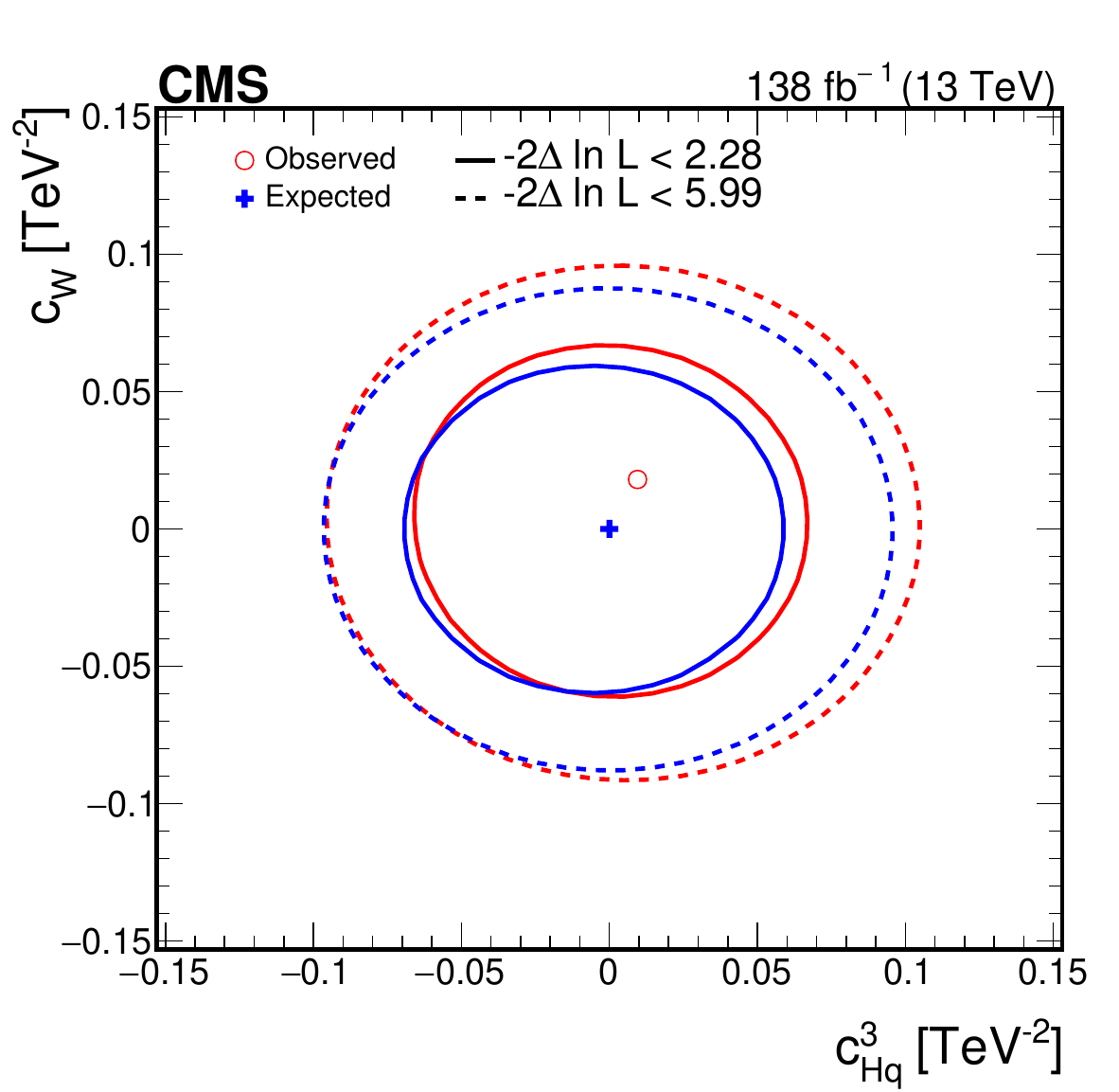} \\
\includegraphics[width=0.45\textwidth]{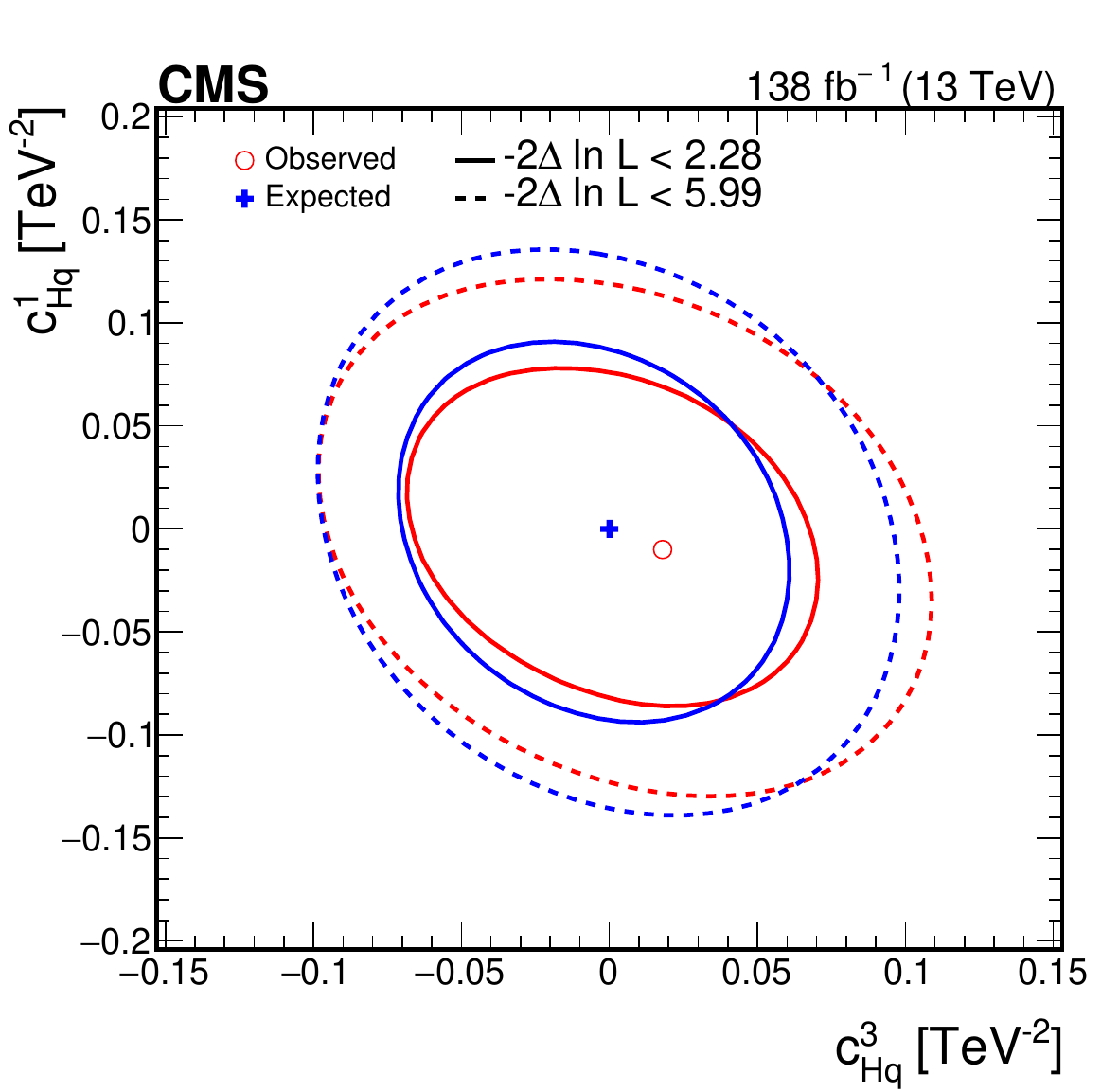}
\includegraphics[width=0.45\textwidth]{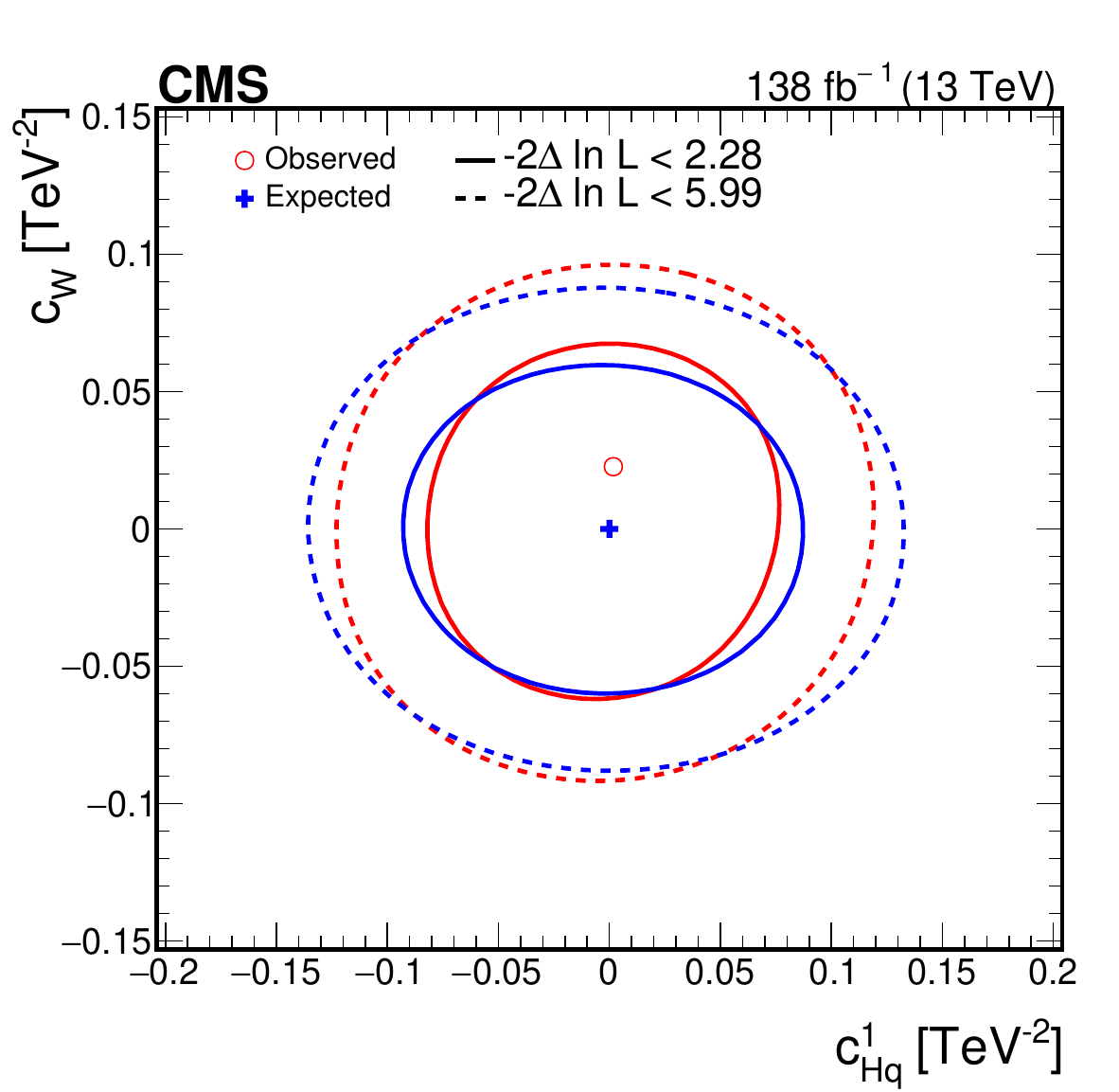}
\caption{Likelihood scans as a function of the pairs of WCs from SMEFT model: $\cjj{3}{8}$ and $\cjj{3}{1}$ (\cmsLeft left), $\chj{3}$ and $\cw$ (\cmsLeft right), $\chj{3}$ and $\chj{1}$ (\cmsRight left), and $\chj{1}$ and $\cw$ (\cmsRight right). All WCs that are not scanned are fixed to zero. The best fit value is shown with a marker and the colored lines correspond to the crossing points of $-2\Delta\ln L$ at 2.28 and 5.99.}\label{fig:2Dlimits_smeft}
\end{figure*}

Figure~\ref{fig:2Dlimits_smeft} shows two-dimensional likelihood scans for pairs of WCs from SMEFT for a selected set of WCs based on the best sensitivity and the correlations between them.
The $\cjj{3}{1}$ and $\cjj{3}{8}$ WCs show a strong linear correlations stemming from their identical Lorentz and $\mathrm{SU}(2)_L$ structure, and differ only by their $\mathrm{SU}(3)$ color structure (singlet versus octet). The $\mwv$ distribution, an inclusive kinematic observable of the $\WV$ system, is not sensitive to the underlying color flow between the quarks produced from $\PV$ boson decays.

\ifthenelse{\boolean{cms@external}}{\clearpage}{}

\section{Summary}

A search for deviations from the standard model  using an effective field theory approach has been presented in the diboson ($\WW$ and $\WZ$) production processes with one W boson decaying to lepton plus antineutrino or charge conjugate and the second $\PW$ or $\PZ$ boson decaying hadronically.
The results are based on data recorded in proton-proton collisions at $\sqrt{s} = 13\TeV$ with the CMS detector at the CERN LHC, corresponding to an integrated luminosity of \lint. 
In this study we constrain Wilson coefficients corresponding to dimension-six effective field theory operators that would lead to anomalous gauge boson self- and vector boson to quark couplings.
Since the contribution from anomalous couplings is expected to be most visible at high energy scales, we focus on final states where the hadronic decay products from the $\PW$ and $\PZ$ bosons merge into a single large-radius jet.
A dedicated classifier, based on machine learning, is employed to separate such jets from vector boson decays from those originating in background processes.
We report the most stringent constraints to date on the Wilson coefficients of operators corresponding to anomalous triple gauge boson couplings.
The bounds set on vector boson to quark couplings are competitive with those from previous inclusive jet measurements.

\begin{acknowledgments}
We congratulate our colleagues in the CERN accelerator departments for the excellent performance of the LHC and thank the technical and administrative staffs at CERN and at other CMS institutes for their contributions to the success of the CMS effort. In addition, we gratefully acknowledge the computing centers and personnel of the Worldwide LHC Computing Grid and other centers for delivering so effectively the computing infrastructure essential to our analyses. Finally, we acknowledge the enduring support for the construction and operation of the LHC, the CMS detector, and the supporting computing infrastructure provided by the following funding agencies: SC (Armenia), BMFWF and FWF (Austria); FNRS and FWO (Belgium); CNPq, CAPES, FAPERJ, FAPERGS, and FAPESP (Brazil); MES and BNSF (Bulgaria); CERN; CAS, MoST, and NSFC (China); MINCIENCIAS (Colombia); MSES and CSF (Croatia); RIF (Cyprus); SENESCYT (Ecuador); ERC PRG and PSG, TARISTU24-TK10 and MoER TK202 (Estonia); Academy of Finland, MEC, and HIP (Finland); CEA and CNRS/IN2P3 (France); SRNSF (Georgia); BMFTR, DFG, and HGF (Germany); GSRI (Greece); MATE and NKFIH (Hungary); DAE and DST (India); IPM (Iran); SFI (Ireland); INFN (Italy); MSIT and NRF (Republic of Korea); MES (Latvia); LMTLT (Lithuania); MOE and UM (Malaysia); BUAP, CINVESTAV, CONACYT, LNS, SEP, and UASLP-FAI (Mexico); MOS (Montenegro); MBIE (New Zealand); PAEC (Pakistan); MSHE, NSC, and NAWA (Poland); FCT (Portugal); J MESTD (Serbia); MICIU/AEI and PCTI (Spain); MOSTR (Sri Lanka); Swiss Funding Agencies (Switzerland); MST (Taipei); MHESI (Thailand); TUBITAK and TENMAK (T\"{u}rkiye); NASU (Ukraine); STFC (United Kingdom); DOE and NSF (USA).

\begin{sloppypar}
\setlength\emergencystretch{\hsize}
\hyphenation{Rachada-pisek} Individuals have received support from the Marie-Curie program and the European Research Council and Horizon 2020 Grant, contract Nos.\ 675440, 724704, 752730, 758316, 765710, 824093, 101115353, 101002207, 101001205, and COST Action CA16108 (European Union); the Leventis Foundation; the Alfred P.\ Sloan Foundation; the Alexander von Humboldt Foundation; the Science Committee, project no. 22rl-037 (Armenia); the Fonds pour la Formation \`a la Recherche dans l'Industrie et dans l'Agriculture (FRIA) and Fonds voor Wetenschappelijk Onderzoek contract No. 1228724N (Belgium); the Beijing Municipal Science \& Technology Commission, No. Z191100007219010, the Fundamental Research Funds for the Central Universities, the Ministry of Science and Technology of China under Grant No. 2023YFA1605804, the Natural Science Foundation of China under Grant No. 12535004, and USTC Research Funds of the Double First-Class Initiative No.\ YD2030002017 (China); the Ministry of Education, Youth and Sports (MEYS) of the Czech Republic; the Shota Rustaveli National Science Foundation (Georgia); the Deutsche Forschungsgemeinschaft (DFG), among others, under Germany's Excellence Strategy -- EXC 2121 ``Quantum Universe" -- 390833306, and under project number 400140256 - GRK2497; the Hellenic Foundation for Research and Innovation (HFRI), Project Number 2288 (Greece); the Hungarian Academy of Sciences, the New National Excellence Program - \'UNKP, the NKFIH research grants K 131991, K 138136, K 143460, K 143477, K 147557, K 146913, K 146914, K 147048, TKP2021-NKTA-64, and 2025-1.1.5-NEMZ\_KI-2025-00004, and MATE KKP and KKPCs Research Excellence and Flagship Research Groups grants (Hungary); the Council of Science and Industrial Research, India; ICSC -- National Research Center for High Performance Computing, Big Data and Quantum Computing, FAIR -- Future Artificial Intelligence Research, and CUP I53D23001070006 (Mission 4 Component 1), funded by the NextGenerationEU program, the Italian Ministry of University and Research (MUR) under Bando PRIN 2022 -- CUP I53C24002390006, PRIN PRIMULA 2022RBYK7T (Italy); the Latvian Council of Science; the Ministry of Science and Higher Education, project no. 2022/WK/14, and the National Science Center, contracts Opus 2021/41/B/ST2/01369, 2021/43/B/ST2/01552, 2023/49/B/ST2/03273, and the NAWA contract BPN/PPO/2021/1/00011 (Poland); the Funda\c{c}\~ao para a Ci\^encia e a Tecnologia (Portugal); the National Priorities Research Program by Qatar National Research Fund; MICIU/AEI/10.13039/501100011033, ERDF/EU, ``European Union NextGenerationEU/PRTR", projects PID2022-142604OB-C21, PID2022-139519OB-C21, PID2023-147706NB-I00, PID2023-148896NB-I00, PID2023-146983NB-I00, PID2023-147115NB-I00, PID2023-148418NB-C41, PID2023-148418NB-C42, PID2023-148418NB-C43, PID2023-148418NB-C44, PID2024-158190NB-C22, RYC2021-033305-I, RYC2024-048719-I, CNS2023-144781, CNS2024-154769 and Plan de Ciencia, Tecnolog{\'i}a e Innovaci{\'o}n de Asturias, Spain; the Chulalongkorn Academic into Its 2nd Century Project Advancement Project, the National Science, Research and Innovation Fund program IND\_FF\_68\_369\_2300\_097, and the Program Management Unit for Human Resources \& Institutional Development, Research and Innovation, grant B39G680009 (Thailand); the Eric \& Wendy Schmidt Fund for Strategic Innovation through the CERN Next Generation Triggers project under grant agreement number SIF-2023-004; the Kavli Foundation; the Nvidia Corporation; the SuperMicro Corporation; the Welch Foundation, contract C-1845; and the Weston Havens Foundation (USA).
\end{sloppypar}
\end{acknowledgments}\section*{Data availability} Release and preservation of data used by the CMS Collaboration as the basis for publications is guided by the  \href{https://doi.org/10.7483/OPENDATA.CMS.1BNU.8V1W}{CMS data preservation, re-use and open access policy}.

\bibliography{auto_generated}

\cleardoublepage \appendix\section{The CMS Collaboration \label{app:collab}}\begin{sloppypar}\hyphenpenalty=5000\widowpenalty=500\clubpenalty=5000\input{SMP-25-001-public-authorlist.tex}\end{sloppypar}
\end{document}

%% file: SMP-25-001-public-authorlist.tex
\cmsinstitute{Yerevan Physics Institute, Yerevan, Armenia}
{\tolerance=6000
A.~Gevorgyan\cmsorcid{0000-0003-2751-9489}, A.~Hayrapetyan, V.~Makarenko\cmsorcid{0000-0002-8406-8605}, A.~Tumasyan\cmsAuthorMark{1}\cmsorcid{0009-0000-0684-6742}
\par}
\cmsinstitute{Institut f\"{u}r Hochenergiephysik, Vienna, Austria}
{\tolerance=6000
P.S.~Hussain\cmsorcid{0000-0002-4825-5278}, M.~Sonawane\cmsorcid{0000-0003-0510-7010}
\par}
\cmsinstitute{Marietta Blau Institute for Particle Physics, Vienna, Austria}
{\tolerance=6000
W.~Adam\cmsorcid{0000-0001-9099-4341}, L.~Benato\cmsorcid{0000-0001-5135-7489}, T.~Bergauer\cmsorcid{0000-0002-5786-0293}, M.~Dragicevic\cmsorcid{0000-0003-1967-6783}, S.~Gundacker\cmsorcid{0000-0003-2087-3266}, A.K.~Guven\cmsorcid{0009-0004-5670-5138}, M.~Jeitler\cmsAuthorMark{2}\cmsorcid{0000-0002-5141-9560}, N.~Krammer\cmsorcid{0000-0002-0548-0985}, A.~Li\cmsorcid{0000-0002-4547-116X}, D.~Liko\cmsorcid{0000-0002-3380-473X}, M.~Matthewman, A.~Pfeiffer\cmsorcid{0000-0001-5328-448X}, J.~Schieck\cmsAuthorMark{2}\cmsorcid{0000-0002-1058-8093}, R.~Sch\"{o}fbeck\cmsAuthorMark{2}\cmsorcid{0000-0002-2332-8784}, M.~Shooshtari\cmsorcid{0009-0004-8882-4887}, N.~Van~Den~Bossche\cmsorcid{0000-0003-2973-4991}, W.~Waltenberger\cmsorcid{0000-0002-6215-7228}, C.E.~Wulz\cmsAuthorMark{2}\cmsorcid{0000-0001-9226-5812}
% ...
\par}
$^{1}$Also at Yerevan State University, Yerevan, Armenia\\
$^{2}$Also at Technische Universit\"{a}t Wien, Vienna, Austria\\
% ...
